\documentclass[12pt,preprint]{aastex}

\slugcomment{Accepted by ApJ December 14, 2010}

\shorttitle{Formation of collapsing cores}
\shortauthors{Kudoh \& Basu}

\begin{document}


\title{Formation of collapsing cores in subcritical magnetic clouds:
three-dimensional MHD simulations with ambipolar diffusion}


\author{Takahiro Kudoh\altaffilmark{1} and Shantanu Basu\altaffilmark{2}}



\altaffiltext{1}{{Division of Theoretical Astronomy, National Astronomical Observatory,
2-21-1 Osawa, Mitaka, Tokyo 181-8588, Japan}
}
\altaffiltext{2}{Department of Physics and Astronomy, University of Western Ontario, 
London, Ontario N6A 3K7, Canada}


\begin{abstract}
We employ the three-dimensional magnetohydrodynamic simulation including 
ambipolar diffusion to study the gravitationally-driven fragmentation of 
subcritical molecular clouds, in which the gravitational
fragmentation is stabilized as long as magnetic flux-freezing applies. 
The simulations show that the cores in an initially subcritical cloud generally 
develop gradually over an ambipolar diffusion time, which is about 
a few $\times 10^7$ 
years in a typical molecular cloud. On the other hand, the formation of collapsing 
cores in subcritical clouds is accelerated by supersonic nonlinear flows. 
Our parameter study demonstrates that core formation occurs faster 
as the strength of the initial flow speed in the cloud increases. 
We found that the core formation time is roughly proportional to the inverse of 
the square root of the enhanced density created by the supersonic nonlinear flows. 
The density dependence is similar to that derived in quasistatically contracting magnetically
supported clouds, although the core formation conditions are created by 
the nonlinear flows in our simulations.
We have also found that the accelerated formation time is not strongly dependent 
on the initial strength of the magnetic field if the cloud is highly subcritical. 
Our simulation shows that the core formation time in our model subcritical clouds 
is several $\times 10^6$ years, due to the presence of large-scale supersonic 
flows ($\sim 3$ times sound speed). 
Once a collapsing core forms, the density, velocity, and magnetic field structure 
of the core does not strongly depend on the initial strength of the velocity fluctuation.
The infall velocities of the cores are subsonic and the magnetic field lines show 
weak hourglass shapes. 

\end{abstract}





\section{Introduction}

Magnetic fields in molecular clouds play an important role in the
early stage of star formation.
Particularly, magnetic fields in molecular clouds can regulate the cloud collapse 
and fragmentation process.
The important parameter for these processes is the mass-to-flux ratio
$M/\Phi$, where $M$ is the mass and $\Phi$ is the magnetic flux of the cloud.
The  mass-to-flux ratio represents the relative strength of gravity and the magnetic field.
There exists a critical mass-to-flux ratio $(M/\Phi)_{\rm crit}$ 
\citep{mes56,str66,mou76,tom88}. 
If $M/\Phi > (M/\Phi)_{\rm crit}$, the cloud is
supercritical and is prone to collapse.
On the other hand, if $M/\Phi < (M/\Phi)_{\rm crit}$, a cloud is
subcritical and cannot collapse as long as magnetic flux-freezing applies.
A similar condition $M/\Phi <  (M/\Phi)_{\rm crit} = 1/(2\pi G^{1/2})$ is required for 
stability against fragmentation of an infinite uniform layer that is flattened along the 
direction of a background magnetic field \citep{nak78}, where $G$ is the gravitational
constant.

\citet{mes56} pointed out that even if clouds are magnetically supported,
ambipolar diffusion will cause the support to be lost and collapse will begin.
More generally, a subcritical cloud undergoes a 
gravitationally-driven fragmentation instability that
occurs on the ambipolar diffusion timescale rather than the dynamical timescale
\citep{lan78,zwe98,cio06}.
The lengthscale of the instability is fundamentally the Jeans-scale in the limit
of highly supercritical clouds, but can be much larger when the mass-to-flux ratio
is close to the critical value \citep{cio06}.
The idea that the star formation is regulated by ambipolar diffusion 
and the magnetic field has been considered for many years
\citep[e.g.,][]{shu87,shu99,mou99}.

Magnetic field strength measurements through the Zeeman effect reveal that 
the mass-to-flux ratios of cloud cores are close to the critical value \citep{cru04}.
These observations are consistent with core formation 
driven by ambipolar diffusion in subcritical clouds.
However, in order to assess whether the fragmentation is occurring in 
a subcritical or supercritical molecular cloud,
magnetic field measurements of the envelope or diffuse region are 
important. \citet{cru09} recently attempted to do this using the Zeeman effect,
and argued that the mass-to-flux ratio 
of the envelope (in four measured positions with beam diameters in the
range of $\sim 0.5 - 1$ pc in the vicinity of four different cloud cores) 
is typically greater than that of the core. This contradicts
the model of core formation in subcritical clouds.
However, \citet{mou09} contested their statistical analysis, and  
argued that dropping the restrictive assumption that the magnetic
field is constant across four measured envelope regions of differing 
morphology and density would result in the opposite conclusion, i.e., that the 
core mass-to-flux ratio is likely greater than that of the envelope.
Future observational tests of this kind, but benefiting from increased amounts of 
source integration time
and spatial resolution,
can settle the current difficulties of interpretation
\citep{cru10}.
The complementary method of 
measurements of polarized emission from dust grains, which reveal
the magnetic field morphology in the cloud, generally show that the 
magnetic field in cloud cores is well ordered, and application of
the Chandrasekhar-Fermi method yields mass-to-flux ratios that are 
also near the critical value 
\citep[e.g.,][]{sch98,gir06}.
\citet{li09} recently compared the magnetic field directions on core scales ($< 1$pc)
with those on large scales ($> 200$ pc) for several molecular clouds, 
and found a significant correlations. 
\citet{alv08} distinctly shows that the magnetic field is locally perpendicular 
to the large filamentary structure of the Pipe Nebula. 
These recent observations of polarized emission indicate that the 
magnetic field provides a dominant force, and that core formation may have 
been driven by ambipolar diffusion in subcritical clouds.

Most nonlinear calculations of ambipolar-diffusion-driven evolution 
in subcritical clouds have focused on a single axisymmetric core,
but some recent models focus on a fragmentation process that results in multiple cores.
Nonlinear calculations of ambipolar-diffusion-driven fragmentation 
in subcritical clouds were first performed by \citet{ind00}, 
who carried out a two-dimensional simulation of an infinitesimally 
thin sheet threaded by an initially perpendicular magnetic field.
\citet{bas04} and \citet{bas09a} carried out two-dimensional simulations of a magnetized sheet
in the thin-disk approximation, which incorporates a finite disk half-thickness
consistent with hydrostatic equilibrium and thereby includes the effect 
of magnetic pressure. 
They found that the fragment spacing in the nonlinear phase agrees with the prediction
of linear theory \citep{cio06}, and that the the subcritical (or critical) model had 
subsonic infall, while the supercritical model had supersonic infall speed. 
The first three-dimensional simulation of the gravitational fragmentation
with magnetic fields and ambipolar diffusion was performed by \citet{kud07},
which verified some of main results of the thin-disk models: 
the dichotomy between subsonic infall speeds in subcritical clouds
and somewhat supersonic speeds in supercritical clouds, for example.

Besides the magnetic field, the supersonic turbulence in molecular
clouds is also an important component in the early stage of star formation 
\citep{mac04,mck07}.
The inclusion of supersonic turbulent initial conditions to a fragmentation
model in subcritical clouds was studied by \citet{li04} and
\citet{nak05}, adopting the thin-disk approximation.
They found that a mildly subcritical cloud can undergo locally rapid 
ambipolar diffusion and form multiple fragments because of the initial supersonic motion
in which the large scale wave mode dominates the power spectrum.
The core formation occurs on the order of turbulence crossing time over the simulation
box, which is comparable to the dynamical time scale.
\citet{bas09b} carried out a parameter study of the fragmentation
regulated by gravity, magnetic fields, ambipolar diffusion, and nonlinear turbulent flows.
They confirmed the onset of runaway collapse in subcritical cloud is significantly
accelerated by the initial nonlinear flows.
\citet{kud08} also verified that the mode of turbulence-accelerated magnetically-regulated star
formation occurs in a fully three-dimensional simulation.
\citet{nak08} also carried out three-dimensional MHD simulations of subcritical clouds 
including star star formation and bipolar outflows, and applied their model to the Taurus molecular cloud.
Since three-dimensional simulations of subcritical clouds are resource-limited, 
large parameter studies have not yet been performed.

In this paper, we study the fragmentation process in subcritical clouds,
including ambipolar diffusion, by fully three-dimensional simulations. 
Our study follows the previous ones of \citet{kud07} and \citet{kud08}.
We focus on the early stage of core formation and evolution, and do not
follow evolution past the runaway collapse of a core.
We carry out a parameter study by running a large number of models, and
with higher spatial resolution than in our previous studies. We are especially interested 
in the effect of the large-amplitude nonlinear initial 
perturbations on the time evolution of the cloud fragmentation, and discuss the mechanism
of turbulence-accelerated star formation in subcritical clouds.

Our paper is organized in the following manner. The numerical model is
described in Section 2, results are given in Section 3, and a discussion
of the results is given in Section 4. We summarize our results in Section 5.

\section{Numerical model}

The numerical model used in this paper is almost the same as that used by 
\citet{kud07} and \citet{kud08}.
We solve the three-dimensional magnetohydrodynamic (MHD) equations including 
self-gravity and ambipolar diffusion.
As an initial condition, we assume hydrostatic equilibrium of 
a self-gravitating cloud along the direction of an initially uniform magnetic field.
We also assume an initially uniform structure perpendicular to the magnetic field lines.
In this equilibrium sheet-like gas, we input a random velocity fluctuation
at each grid point.

\subsection{Basic Equations}
\label{equation}

We solve the three-dimensional MHD equations including self-gravity and ambipolar diffusion,
assuming that neutrals are much more numerous than ions:

\begin{equation}
\frac{\partial \rho}{\partial t} 
+ \mbox{\boldmath$v$} \cdot \nabla \rho
= -\rho \nabla \cdot \mbox{\boldmath$v$},
\label{eq:continuity}
\end{equation}

\begin{equation}
\frac{\partial \mbox{\boldmath$v$}}{\partial t} 
+ (\mbox{\boldmath$v$} \cdot \nabla) \mbox{\boldmath$v$}
=-\frac{1}{\rho} \nabla p
+ \frac{1}{c \rho} \mbox{\boldmath$j$} \times \mbox{\boldmath$B$}
- \nabla \psi,
\label{eq:momentum}
\end{equation}

\begin{equation}
\frac{\partial \mbox{\boldmath$B$}}{\partial t}
= \nabla \times (\mbox{\boldmath$v$} \times \mbox{\boldmath$B$})
+ \nabla \times \left[\frac{\tau_{ni}}{c\rho} (\mbox{\boldmath$j$} \times \mbox{\boldmath$B$}) \times \mbox{\boldmath$B$}\right],
\label{eq:induction}
\end{equation}

\begin{equation}
\mbox{\boldmath$j$}=\frac{c}{4\pi} \nabla \times \mbox{\boldmath$B$},
\end{equation}

\begin{equation}
\nabla ^2 \psi = 4\pi G \rho,
\label{eq:poisson}
\end{equation}

\begin{equation}
p=c_s^2 \rho,
\end{equation}
where $\rho$ is the density of neutral gas, $p$ is the pressure, 
{\boldmath$v$} is the velocity, {\boldmath$B$} is the magnetic field,
{\boldmath$j$} is the electric current density, 
$\psi$ is the self-gravitating potential, 
and $c_s$ is the sound speed.
Here, we consider the molecular clouds whose number density 
is about $10^3 - 10^6$ cm$^{-3}$. In this case, the cooling time
is generally much smaller than the dynamical time. Therefore,
%
instead of solving a detailed energy equation, we assume isothermality
for each Lagrangian fluid particle 
\citep{kud03,kud06}:
\begin{equation}
\frac{d c_s}{dt} 
= \frac{\partial c_s}{\partial t} 
+ \mbox{\boldmath$v$} \cdot \nabla c_s = 0.
\label{eq:isothermal}
\end{equation}
This means that each parcel of molecular cloud gas as well as surrounding
warm gas (see Sec. \ref{initcond}) retains its initial temperature.
For the neutral-ion collision time in equation (\ref{eq:induction})
and associated quantities,
we follow \citet{bas94}, so that
\begin{equation}
\tau_{\rm ni} =  1.4\, \frac{m_i+m_n}{\rho_i \langle \sigma w \rangle_{in}},
\label{eq:tau}
\end{equation}
where $\rho_i$ is the density of ions and $\langle\sigma w\rangle_{\rm in}$ 
is the average collisional rate between ions of mass $m_i$ and neutrals
of mass $m_n$. Here, we use typical values of HCO$^+$-H$_2$ collisions, 
for which $\langle\sigma w\rangle_{\rm in}=1.69 \times 10^{-9}$ cm$^{-3}$s$^{-1}$ 
, $m_{\rm i}/m_{\rm n} =14.4$, and $m_{\rm n}=2 \times 1.66 \times 10^{-24}$ g. We also assume that
the ion density $\rho_i$ is determined by the approximate relation
\citep{elm79,nak79}
\begin{equation}
\rho_{\rm i}=m_{\rm i} K \left(\frac{\rho/m_{\rm n}}{10^5\, \mbox{cm$^{-3}$}}\right)^k,
\label{eq:roh_i}
\end{equation}
where we assume $k=0.5$ throughout this paper, and take 
$K=3 \times 10^{-3}$cm$^{-3}$ as a typical value.
By using equation (\ref{eq:roh_i}), equation (\ref{eq:tau}) can be
written as
\begin{equation}
\tau_{\rm ni}=\gamma \rho^{-1/2} ,
\label{eq:tni}
\end{equation}
where $\gamma \simeq 170.2$ g$^{1/2}$cm$^{-3/2}$ s using the above typical values.
Then, equation (\ref{eq:induction}) becomes
\begin{equation}
\frac{\partial \mbox{\boldmath$B$}}{\partial t}
= \nabla \times (\mbox{\boldmath$v$} \times \mbox{\boldmath$B$})
+ \nabla \times \left[\frac{\gamma}{c\rho^{3/2}} (\mbox{\boldmath$j$} \times \mbox{\boldmath$B$}) 
\times \mbox{\boldmath$B$}\right] .
\label{eq:induction2}
\end{equation}

\subsection{Initial Conditions}
\label{initcond}

As an initial condition, we assume hydrostatic equilibrium of 
a self-gravitating one-dimensional cloud along the $z$-direction
\citep{kud03,kud06}. The hydrostatic equilibrium is calculated 
from the equations

\begin{equation}
\frac{dp}{dz}=\rho g_z,
\label{eq:hsp}
\end{equation}

\begin{equation}
\frac{dg_z}{dz}=-4\pi G \rho,
\label{eq:hsg}
\end{equation}

\begin{equation}
p=c_s^2 \rho,
\label{eq:hss}
\end{equation}
subject to the boundary conditions
\begin{equation}
g_z(z=0)=0,\ \ \rho(z=0)=\rho_0,\ \ p(z=0)=\rho_0\, c_{s0}^2
\end{equation}
where $\rho_0$ and $c_{s0}$ are the initial density and
sound speed at $z=0$.
If the initial sound speed (temperature) is uniform throughout the region,
we have the following analytic solution $\rho_S$ found by \citet{spi42}:
\begin{equation}
\rho_S(z)=\rho_0 \,\mbox{sech}^2(z/H_0),
\end{equation}
where
\begin{equation}
H_0=\frac{c_{s0}}{\sqrt{2\pi G \rho_0}}
\label{eq:h0}
\end{equation}
is the scale height.
However, an isothermal molecular cloud is usually surrounded
by warm material, such as neutral hydrogen gas.
Observations also show that the transition between molecular gas and
surrounding gas is quite sharp \citep{bli91}.
In order to simulate the situation,
we assume the initial sound speed distribution to be
\begin{equation}
c_{s}^2(z)=c_{s0}^2 
+ \frac{1}{2} (c_{sc}^2 - c_{s0}^2)
\left[ 1+\tanh \left(\frac{|z|-z_c}{z_d}\right) \right] ,
\end{equation}
where we take $c_{sc}^2=10\,c_{s0}^2$, $z_c=2H_0$, and $z_d=0.1H_0$
throughout the paper. 
This equation gives a numerical model of the temperature transition of 
$c_{sc}^2/c_{s0}^2$ at $z_c$ with the transition length of $z_d$.
By using this sound speed distribution, we can solve equations
(\ref{eq:hsp})-(\ref{eq:hss}) numerically. The initial density distribution 
of the numerical solution shows that it is almost the same as Spitzer's 
solution for $0 \leq z \leq z_c$.

We also assume that the initial magnetic field is uniform along the $z$-direction:
\begin{equation}
B_z=B_0,\ \ B_x=B_y=0,
\end{equation}
where $B_0$ is constant.

In this equilibrium sheet-like gas, we input a random velocity fluctuation
at each grid point:
\begin{equation}
v_x=v_a R_m(x,y),\ \ v_y=v_a R_m(x,y),\ \ v_z=0.0,
\end{equation}
where $R_m$ is a random number with spectrum $v_k^2 \propto k^{n}$ in Fourier space,
where $k=(k_x^2+k_y^2)^{1/2}$, and the root mean square value is about 1. 
The $R_m$'s for each of $v_x$ and $v_y$ are independent realizations.
In this paper, we apply the two distinct spectra of $v_k^2 \propto k^{4}$ and $v_k^2 \propto k^{0}$. 
The parameter $v_a$ shows the strength of the velocity fluctuation.
Models with the same spectrum use the same pair of realizations 
of $R_m$ for generating the initial perturbations.

\subsection{Numerical Parameters}

The constant $G$ along with the two parameters $c_{s0}$ and $\rho_0$ 
that are associated with the initial state allow one to choose
a set of three fundamental units for this problem. We choose these
to be $c_{s0}$, $H_0$, and $\rho_0$ for velocity, length, and density,
respectively.
These yield a time unit $t_0 \equiv H_0/c_{s0}$.
The initial magnetic field
strength introduces one dimensionless free parameter,
\begin{equation}
\beta_{0} \equiv \frac{8 \pi p_0}{B_0^2} 
= \frac{8 \pi \rho_0 c_{s0}^2}{B_0^2},
\end{equation}
the initial ratio of gas to magnetic pressure 
at $z=0$.

In the sheet-like equilibrium cloud with a vertical 
magnetic field, $\beta_{0}$ is related to the mass-to-flux ratio
for Spitzer's self-gravitating cloud. The mass-to-flux ratio 
normalized to the critical value is 
\begin{equation}
\mu_S \equiv 2\pi G^{1/2} \frac{\Sigma_S}{B_0} 
\end{equation}
where
\begin{equation}
\Sigma_S=\int_{-\infty}^{\infty} \rho_S dz = 2\rho_0 H_0
\end{equation}
is the column density of Spitzer's self-gravitating cloud.
Therefore,
\begin{equation}
\beta_{0}=\mu_S^2.
\end{equation}
Although the initial cloud we used is not exactly 
the same as the Spitzer cloud, $\beta_{0}$ is a good 
indicator to whether or not the magnetic field 
can prevent gravitational instability
in the flux-freezing case
\citep{nak78}.

We also define a dimensionless parameter
\begin{equation}
\alpha \equiv \gamma \sqrt{2 \pi G},  
\end{equation}
which shows the strength of the ambipolar diffusion.
We note that $\alpha$ can also be equated to $\tau_{\rm ni,0}/t_0$, where 
$\tau_{\rm ni,0}$ is the initial value of the neutral-ion collision time
calculated from equation (\ref{eq:tni}).
By using the typical values in Section 2.1, we get $\alpha \simeq 0.11$, 
which is taken as a fiducial parameter in this paper. 
We also vary $\alpha$ as a free parameter in Section 3.3.


Dimensional values of all quantities can be found 
through a choice of $\rho_0$ and $c_{s0}$.
For example, for $c_{s0}=0.2$ km s$^{-1}$ and $n_0=\rho_0/m_n=10^4$ cm$^{-3}$, 
we get $H_0 \simeq 0.05$ pc, $t_0 \simeq 2.5 \times 10^5$ yr, and 
$B_0 \simeq 20\,\mu$G if $\beta_0=1$.

\subsection{Numerical Technique}

In order to solve the equations numerically, we use the 
CIP method \citep{yab91a,yab01}
for equations (\ref{eq:continuity}), (\ref{eq:momentum}) 
and (\ref{eq:isothermal}), and the method of 
characteristics-constrained transport (MOCCT; Stone \& Norman 1992) 
for equation (\ref{eq:induction}), including an explicit integration
of the ambipolar diffusion term. The combination of the CIP and 
MOCCT methods is summarized in \citet{kud99} and
\citet{oga04}. It includes the CCUP method 
\citep{yab91b} for the calculation of gas pressure, 
in order to get more numerically stable results.
The numerical code used for this paper is based on 
that of \citet{oga04}. 

In this paper, the ambipolar diffusion term is only included when the
density is greater than a certain value, $\rho_{cr}$.
We let $\rho_{cr}=0.3\rho_0$ both for numerical convenience
and due to the physical idea 
that the low density region is affected 
by external ultraviolet radiation so that the ionization fraction
becomes large, i.e., $\tau_{\rm ni}$ becomes small
\citep{cio95}.
Under this assumption, the upper atmosphere of the sheet-like cloud
is not affected by ambipolar diffusion.
This simple assumption helps to avoid very small time-steps
due to the low density region in order to maintain stability of
the explicit numerical scheme. 

We use a mirror-symmetric boundary 
condition at $z=0$ and periodic boundaries in the $x$- 
and $y$-directions. At the upper boundary, 
at $z=z_{\rm out}=4H_0$, we also use a mirror-symmetric boundary 
except when we solve for the gravitational potential. This symmetric condition 
is just for numerical convenience. 
When the same boundary conditions were used by \citet{kud07} and \citet{kud08}
for three-dimensional simulations, the results were consistent 
with two-dimensional  thin-disk simulations 
\citep[e.g.,][]{bas04,bas09b}.
Hence, we believe that the boundary conditions 
do not affect the result significantly. 
The Poisson equation (\ref{eq:poisson}) 
is solved by the Green's function method to compute the gravitational 
kernels in the $z$-direction, along with a Fourier transform 
method in the $x$- and $y$-directions \citep{miy87b}.
This method of solving the Poisson equation allows us to 
find the gravitational potential of a vertically isolated cloud 
within $|z|<z_{\rm out}$.

The computational region is $|x|,|y| \leq 8\pi H_0$ and 
$0 \leq z \leq 4H_0$. 
The number of grid points for each direction is 
$(N_x,N_y,N_z)=(256,256,40)$.
Since the most unstable wavelength for no magnetic field is about 
$4\pi H_0$ 
\citep{miy87a},
we have 64 grid points within this wavelength.
We have also 10 grid points within the scale height of the initial cloud
in the $z$-direction. 
The computational time of the fiducial model is about 70 hours of 
CPU time using four processors of the SX-9 in the National Astronomical 
Observatory of Japan. 
The maximum computational time, which occurs for 
the case of the highly subcritical cloud with small initial velocity fluctuations, 
is about 770 hours of CPU time.
In the cases of the supercritical model, the computational times are about
30 minutes of CPU time.

\section{Results}

Table 1 summarizes the models and parameters for simulations presented in this paper.
In the table, we have listed the values of the free parameters
$\beta_{0}$,  $\alpha$, the form of the turbulent power spectrum, and $v_a$.
We also have listed the core formation time $t_{core}$, which is defined at the time
when the density of the core reaches 100 $\rho_0$. Although the definition 
of $t_{core}$ is determined from a practical constraint of the numerical simulation, 
in practice the center of the core is always supercritical at this time and the time
evolution of the density shows the features of runaway collapse.
In the models V0 to V5, we change the amplitude of the initial velocity fluctuation $v_a$,
with the form of the turbulent power spectrum fixed at $v_k^2 \propto k^{-4}$.
In the models K0 to K4, we change the amplitude of the initial velocity fluctuation,
but with the form of the turbulent power spectrum fixed at $v_k^2 \propto k^{0}$.
In the models A0 to A4, we change the dimensionless ambipolar diffusion coefficient 
$\alpha$ for models with $v_a=3.0c_{s0}$ or $v_a=0.1c_{s0}$.
In the models B0 to B9, we change $\beta_{0}$, the initial plasma beta at $z=0$,
which corresponds to the square of the initial mass-to-flux ratio, 
for models with $v_a=3.0c_{s0}$ or $v_a=0.1c_{s0}$.
We are mostly interested in core formation in subcritical clouds, but 
also carry out some supercritical cases for comparison.

\subsection{General properties of a fiducial model} 


We show the result of model V4 as a fiducial model,
in which the initial turbulent velocity amplitude ($v_a$) is 3 times sound speed ($c_{s0}$), 
its spectrum is $v_k^2 \propto k^{-4}$, the initial normalized mass-to-flux ratio 
is about 0.5 (i.e., $\beta_0=0.25$), and the dimensionless ambipolar diffusion 
coefficient has a typical value ($\alpha=0.11$).
We suppose that these values are approximately typical in molecular clouds.

Figure 1 shows the time snapshot of the logarithmic density colour map overlaid with 
the velocity vector field for model V4. 
The snapshot is at the end of our simulation, when the maximum density is greater
than 100$\rho_0$. The upper panel shows the cross section at $z=0$, and 
the lower panel shows it at $y=20.7H_0$.
The maximum density is located at $(x,y,z)=(-13.4H_0,20.7H_0,0)$.
A collapsing core is located in the vicinity of the maximum density.
Figure 2 shows the time snapshot of the logarithmic plasma $\beta$,
the ratio of gas pressure to magnetic pressure, at the end of the simulation. 
It shows that $\beta$ is greater than 1 around the core.
It means that once the collapsing core is formed, the mass-to-flux ratio of the core
is expected to be greater than 1 ($\beta > 1$), i.e., supercritical.
The other collapsing cores are also formed in the region where plasma $\beta$ is
greater than 1 (the left-down region of the upper panel in Fig. 2).
The left panel of Figure 3 shows the density, $x$-velocity, and plasma $\beta$
along an $x$-axis cut at $y=20.7H_0$ and $z=0$. 
The right panel shows the density, $z$-velocity, and plasma $\beta$ 
along a $z$-axis cut at $x=-13.4H_0$ and $y=20.7H_0$.
In each panel, the velocity shows infall motion 
into the center of the core, though the $x$-velocity outside of the core 
(low density region) is affected by turbulent motions originating
from the initial perturbation. The relative infall speed is subsonic and
is about half of the sound speed at that time.
The subsonic infall feature is consistent with previous studies of core formation
in subcritical clouds \cite[e.g.,][]{cio00,bas04}.

Figure 4 shows the time evolution of the maximum density at $z=0$ 
and $\beta$ at the location of the maximum density.
The density goes up to $\sim 10$ times greater than the initial density 
during the first compression of the cloud. Then, it rebounds and shows
oscillations. Eventually, the dense region goes into runaway collapse around $t=18t_0$,
which corresponds to about $5 \times 10^6$ years.
In the first compression, $\beta$ also goes up to $\sim 0.6$. 
It gradually goes up and becomes greater than $1$ when the runaway 
collapse occurs.
As we discussed in the previous section, $\beta$ roughly equals the square 
of the normalized mass-to-flux ratio in the gravitational equilibrium state
of gas with vertical magnetic field. 
The runaway collaping core is expected to be supercritical ($\beta >1$).
Figure 4, Figure 2, and Figure 3 demonstrate that $\beta$ is a good indicator of the 
mass-to-flux ratio, though the gas is not exactly in gravitational 
equilibrium at the last stage of our simulations. 

Figure 5 shows a close-up view of the core in the vicinity of $(x,y,z)=(-13.4H_0,20.7H_0,0)$.
The iso-surface contour shows the logarithmic density and the lines 
represent the magnetic field.
The core is located in a filamentary structure that was induced by the initial velocity
fluctuation. The magnetic field lines show an hourglass shaped structure because 
of the infall motion into the center of the core.

\subsection{Effect of initial velocity fluctuations} 

Figure 6 shows a time snapshot of the logarithmic density overlaid with velocity vectors 
for model V0. 
The model V0 coresponds to the model for initial velocity amplitude ($v_a$)
is 0.01 times sound speed ($c_{s0}$), but other parameters are the same as 
the fiducial model V4.
The snapshot is at the end of our simulation, when the maximum 
density is greater than 100$\rho_0$. When the initial perturbed velocity is much 
smaller than the sound speed, the core formation time ($t_{core} \sim 136t_0$,
which corresponds to $\sim 3\times 10^7$ years) 
is much longer than that ($t_{core} \sim 18.7t_0$) of model V4. 
The maximum density is located at $(x,y,z)=(7.6H_0,-11.9H_0,0)$.
A collapsing core is formed in the vicinity of the maximum density.
Figure 7 shows the time snapshot of the logarithmic plasma $\beta$ 
at the end of the simulation for model V0. 
As we expected, 
the plasma $\beta$ value is greater
than 1 around the core
,which means the core is supercritical.
In Figure 6 and Figure 7, core-like structures are gathered together 
on the side of $x > 0$. This reflects the initial perturbation of $k^{-4}$ 
in which the larger scale contains greater energy in the velocity spectrum.
The left panel of Figure 8 shows the density, $x$-velocity, and plasma $\beta$
along an $x$-axis cut at $y=7.6H_0$ and $z=0$. 
The right panel shows the density, $z$-velocity, and plasma $\beta$ 
along a $z$-axis cut at $x=-7.6H_0$ and $y=-11.9H_0$.
The velocity shows infall motion into the center of the core, though
the center of the core shows a systematic speed in the positive
$x$-direction of about $0.3\,c_{s0}$. 
The relative infall speed is also subsonic.
Figure 9 shows the close up view of the core in the vicinity of $(x,y,z)=(-13.4H_0,20.7H_0,0)$.
The iso-surface contour shows the logarithmic density and the lines represent the magnetic 
field. The core shows typical oblate-like structure.
The magnetic field line shows a hourglass shaped structure and the deformation
of the field line is similar to that of model V4 in Figure 5.

Figure 10 shows the time snapshot of the logarithmic density overlaid with velocity vectors 
at the end of the simulation for model K1. 
The model K1 coresponds to the model for the initial velocity 
spectrum is $v_k^2 \propto k^{0}$, but other parameters are the same as 
the fiducial model V4.
The flat spectrum is possibly not consistent with observation, but it is used to
compare with the result by the typical spectrum of $v_k^2 \propto k^{-4}$.
When the initial perturbed velocity 
has a flat spectrum, the core formation time ($t_{core} \sim 67t_0$) is longer
than that ($t_{core} \sim 18.7t_0$) of model V4.
The maximum density is located at $(x,y,z)=(-1.5H_0,-9.5H_0,0)$ and a collapsing core 
is formed in the vicinity of the maximum density.
Figure 11 shows the time snapshot of the logarithmic plasma $\beta$ 
at the end of the simulation for model K1. 
We note that the $\beta$ is greater than 1 around the core again.
The left panel of Figure 12 shows the density, $x$-velocity, and plasma $\beta$
along an $x$-axis cut at $y=-9.5H_0$ and $z=0$. 
The right panel shows the density, $z$-velocity, and plasma $\beta$ 
along a $z$-axis cut at $x=-1.5H_0$ and $y=-9.5H_0$.
The velocity shows infall motion into the center of the core, 
and the relative infall speed is also subsonic again.
Figure 13 shows a close-up view of the core in the vicinity of $(x,y,z)=(-1.5H_0,-9.5H_0,0)$.
The magnetic field lines also show an hourglass shaped structure.
Figure 3, Figure 8 and Figure 12 show the structure of collapsing cores are not so dependent on the
initial velocity fluctuations, except that the outer region of the core in model V4 is 
strongly affected by the large-scale turbulent flow. 

The main difference of each model is the core formation time.
Figure 14 shows the time evolution of the maximum density for models V4, V0 and K1.
In the case of model V4, which has a $k^{-4}$ spectrum, the maximum density is 
strongly increased by the compression caused by the large scale supersonic flow. 
In the case of model K1, 
the compression occurs in the initial stage, but it is weaker than that of 
model V4 because the small scale perturbations contain more power than 
in model V4. In the case of model V0, with small-amplitude initial perturbations,
the core formation time is about 7 times longer than that of model V4.

Figure 15 shows the core formation time as a function of the strength of the 
initial velocity perturbation. As the perturbed velocity amplitude increases, 
the core formation time decreases for each type of
initial spectrum. In the cases of the initial $k^{-4}$ spectrum (filled squares), 
the core formation times are systematically smaller than those of the
initial $k^{0}$ spectrum (open squares).
In order to understand the physics of the core formation time,
we plot the core formation time as a function of $\rho_{peak}$ in Figure 16, where 
$\rho_{peak}$ is defined as the value of the density peak during the first compression 
in the time evolution of the maximum density on $z=0$.
In the cases of the $k^{-4}$ spectrum (filled squares), $\rho_{peak}$ is greater
than in the cases of the $k^{0}$ spectrum (open squares), even when $v_a$ is relatively small. 
Figure 16 shows that the core formation time is shorter when the density peak 
$\rho_{peak}$ is greater, and indicates that it is 
 nearly proportional to $1/\sqrt{\rho_{peak}}$.
The density dependence is similar to that derived in quasistatically contracting magnetically
supported cores, assuming a force balance of the magnetic force and gravity in the core 
\citep[e.g.,][]{mou99}.
In our simulation, the density is increased by the compression caused by large-scale 
nonlinear flows, and a dense region undergoes a rebound and oscillations, which means the core 
is not in an exact force balance.
Nevertheless, it is interesting that the core formation time is
nearly proportional to $1/\sqrt{\rho_{peak}}$.
We discuss this subject further in Section 4.

\subsection{Dependence on ambipolar diffusion coefficient} 

Figure 17 shows the time evolution of the maximum density for models V4, A0, A1 and A2.
The model A0, A1, and A2 correspond to the models for $\alpha=0$, $\alpha=0.05$,
and $\alpha=0.2$, respectively, but other parameters of them are the same as those of V4.

As $\alpha$ decreases, the core formation time increases. When $\alpha=0$ (model A0), 
which means no ambipolar diffusion, a core was not formed during the calculation time, 
lasting until $t \simeq 268t_0$.
Figure 18 shows the core formation time as a function of $\alpha$. When the initial 
velocity perturbation is small, the core formation time shows nearly linear
dependence on $\alpha^{-1}$ (open triangles).
When the initial perturbation is nonlinear (filled triangles), 
the dependence on $\alpha^{-1}$ is 
a little steeper than a linear dependence.
For model A2 ($\alpha=0.2$, $v_a=3c_{s0}$), the core formation occurs very rapidly, without 
showing an oscillation of the maximum density.
When $\alpha$ is large enough, the supercritical region can appear quickly 
after the first compression, resulting in a significantly reduced core formation time.

In figure 19, we show the time evolution of the total kinetic energies in the 
computational region for model V4, A0, A1, and A2.
The oscillation of the energies comes from the strong compression 
and rebound of the clouds.
The figure shows that the kinetic energies quickly decrease within 
a crossing time of the initial perturbed velocity across the computational region. 
The $e$-folding time of the energies depends slightly on $\alpha$.
As $\alpha$ increases, the $e$-folding time increases.
In the case of $\alpha=0$ (model A0), about 10 percent of the initial kinetic energy 
remains at the end of the calculation ($t \simeq 268t_0$).
Basu \& Dapp (2010) found that the significant portion ($\sim 1/2$) 
of the initial kinetic energy remains in the flux-freezing case using a 
thin-disk approximation.
Our result shows that the kinetic energy is more dissipative than in
the thin-disk approximation.
This is because of the different boundary condition in the vertical direction.
We discuss this subject further in Section 4.

\subsection{Dependence on initial mass-to-flux ratio} 

Figure 20 shows the time evolution of the maximum densities for models V4, B0, B1 and B2,
showing the difference between models with different $\beta_{0}$.
The model B0, B1, and B2 correspond to the models for $\beta=0.04$, $\beta=0.09$, 
and $\beta=0.36$, respectively, but other parameters of them are the same as those 
of V4.
As we have shown in Section 2, $\beta_{0}$
approximately equals the square of the initial normalized mass-to-flux ratio 
of the cloud.
The figure shows that the core formation time is longer when $\beta_{0}$ is smaller.
The figure also shows that the initial density enhancement is smaller when $\beta_{0}$ is smaller.
Figure 21 shows the core formation time as a function of $\beta_{0}$. 
When $\beta_{0}$ is greater than 1, the initial cloud is supercritical.
In the case of the supercritical clouds, 
the core formation occurs in dynamical time scale.
When the initial velocity is large, a core forms during the first compression. 
The core formation time of the supercritical clouds does not strongly depend on
the initial mass-to-flux ratio (or $\beta_{0}$).
When $\beta_{0}$ is smaller than 1, the initial cloud is subcritical.
The core formation times of a subcritical cloud are generally longer
than those of a supercritical cloud.
In the case of a highly subcritical cloud however, 
the core formation time does not strongly depend on the initial mass-to-flux ratio.
The transition occurs near the critical cloud, $\beta_{0} \sim 1$.
This tendency is consistent with that of the growth time from the linear analysis using
the thin-disk approximation \citep{cio06} and with that of the ambipolar diffusion time
in quasistatically contracting magnetically supported cores \citep{mou99}.
It is interesting that the same tendency is achieved even when the
initial velocity fluctuation is nonlinear (filled circles).


\section{Discussion}


\subsection{Timescale of core formation}
One of the problems for core formation in subcritical clouds is that the timescale
of the core formation is slower than that is expected from observations
\citep[e.g.,][]{jij99}.
In order to solve this problem, \cite{zwe02} and \cite{fat02} pointed out that 
the ambipolar diffusion rate can be enhanced in 
a turbulent medium with a fluctuating magnetic field.
In a different way, \citet{li04} and \citet{nak05} found that the compression 
of the cloud by the large-scale turbulent flow shortened the timescale of core formation 
even in subcritical clouds.
\cite{kud08} and this paper followed the latter idea using three-dimensional 
simulations without a thin-disk approximation. 
Here, we discuss how the compression shortens the timescale of core formation 
in subcritical clouds.

Figure 16 indicates that the core formation time is nearly proportional to 
$1/\sqrt{\rho_{peak}}$, where $\rho_{peak}$ 
is the value of the density peak during the first compression in the time evolution
of the maximum density at $z=0$. 
The ambipolar diffusion time ($\tau_{\rm ad}$) is estimated from equation (\ref{eq:induction2})
as $\tau_{\rm ad} \sim 4\pi (\sqrt{2\pi G}/\alpha) \rho^{3/2}L^2/B^2$, 
where $L$ is the gradient length scale
introduced by the turbulent compression. Because the compression by the nonlinear flow
is nearly one-dimensional, the
magnetic field scales roughly as $B \propto L^{-1}$ within
the flux-freezing approximation. The surface density $\Sigma$ also scales 
as $\Sigma \propto L^{-1}$. If the compression is rapid enough that 
vertical hydrostatic equilibrium cannot established, then 
$\rho \propto \Sigma \propto L^{-1}$, and 
$\tau_{ad} \propto L^{5/2} \propto \rho^{-5/2}$ \citep{elm07,kud08}.
This means that diffusion can occur quickly if the turbulent compression creates 
small values of $L$ or large values of $\rho$. This would lead to a rapidly rising
value of $\beta$ in Figure 4 at $t/t_0 \sim 1$.
If diffusion is so effective during the first turbulent compression that a dense
region becomes supercritical, then it will evolve directly into collapse.
Alternatively, the stored magnetic energy of the compressed (and still subcritical) 
region may lead to reexpansion of the dense region. 
If reexpansion of the initial compression does occur, then there is enough time for
the vertical structure settle back to near-hydrostatic equilibrium.
In this case, the density scales roughly like $\rho \propto \Sigma^2 \propto L^{-2}$,
and $\tau_{\rm ad} \propto L \propto \rho^{-1/2}$.
The ambipolar diffusion time is also estimated by \citet{mou99} as
$\tau_{\rm ad} \propto \rho^{-1/2}$ in quasistatically contracting magnetically
supported cores, assuming a radial force balance between gravity
and the magnetic Lorentz force.
In our simulation, force balance is not exactly achieved, but the
time average (of the oscillations) in the cores can be approximately 
in force balance.
Since the first strong compression leads to rapid magnetic flux loss,
a higher density region (than the initial value) is rapidly attained 
and then settles into an approximate force balance.
Since the phase of continuing ambipolar diffusion in the oscillating 
high density region takes longer than the initial compression, the
overall time scale of the core formation is approximately proportional
to $\rho_{\rm peak}^{-1/2}$. 
Here, instead of the time-average density, 
$\rho_{\rm peak}$ is used as a representitive density of the compressed gas, 
since it is complicated to determine the time-average density of the moving, 
oscillating and collapsing high density region. 
It is interesting that the relation attains even when the $\rho_{\rm peak}$ is used. 
We expect that the average density of the compressed overdense region would 
nearly proportional to the peak density. 

The relation means that the density dependence of
$\tau_{\rm ad} \propto \rho^{-1/2}$ is the same as that of the
free fall time. Figure 16 shows that the actual time of the core 
formation is about 30 times longer than the free fall time 
of gas with $\rho_{peak}$ when $\alpha=0.11$ and $\beta_0=0.5$.
The vale of the dimensionless coefficient ($\sim 30$) is expected to be inversely 
proportinal to $\alpha$ (section 3.3), 
and not to be strongly dependent on the mass-to-flux ratio except 
when it is nearly critical (section 3.4).
Even when the time-average density is used instead of the peak density,
the value of the dimensionless coefficient would not be so different 
because of the weak dependence of $\rho$ ($\propto \rho^{-1/2}$).

In some cases, collapse may occur during the first compression itself or
soon thereafter, and the approximate scaling of the core formation time with
$\rho_{\rm peak}^{-1/2}$ will not hold.  
Whether or not this occurs depends not only on
the strength of the turbulent compression, but also on the initial 
mass-to-flux ratio of the cloud (related to $\beta_{0}$) and the ambipolar diffusion 
coefficient ($\alpha$). In model B2 (a subcritical model that is closest to
critical, with normalized initial mass-to-flux ratio $\simeq 0.6$), 
a reexpansion does occur but the collapse starts relatively quickly, 
in the third oscillation (Fig. 20). 
In model A2 ($\alpha=0.2$), which has the poorest neutral-ion coupling of all
models, the collapse starts almost in the first compression (Fig. 17).

An overall conclusion is that
the core formation occurs more rapidly than it would in the initial
state due to the elevated value of $\rho$ in the compressed but oscillating region.
Since the core formation time is approximately proprtional to $\rho^{-1/2}$, 
$10-100$ times desity enhancement is needed to get $3-10$ times shorter core formation 
time than that of the standard core formation model in subcritical clouds
\citep[e.g.][]{jij99}.
The observed non-thermal velocity ($3-10$ times sound speed) is eligible to
make such enhancemenr by the compression in the isothermal clouds, 
if its scale is larger than the Jeans length.

\subsection{Structure of cores}

Even though the core formation time is accelerated by the nonlinear flows,
the density, velocity, and magnetic field structure of a core does not strongly 
depend on the initial strength of the velocity fluctuation
(e.g., Fig. 3, Fig. 8, and Fig. 12).
The infall velocities of the cores are subsonic and the magnetic field lines show 
weak hourglass shapes. 
This result may be consistent with the result that the core formation time is proportional
to $\rho_{\rm peak}^{-1/2}$.
Even when core formation is initiated by the initial turbulence, the core 
properties can be similar to the quasistatically contracting magnetically
supported cores discussed by \citet{mou99}.
The initial turbulence accelerates core formation, but it eventually dissipates 
in the dense region when the collapsing core is formed.
Subsonic infall motions were found by \cite{lee01} in an observational survey of 
starless cores. They found that the typical infall radii are $0.06 - 0.14$ pc
and that the infall speed lies in the range of $0.05 - 0.09$ km s$^{-1}$.
These values are consistent with our results.
The subsonic infall is the common feature of core formation in subcritical clouds,
which has been pointed out by 1D axisymmetric \citep{cio00} 
and 2D thin-disk \citep[]{bas04,bas09a,bas09b} models.
Because of the subcritical infall, the hourglass shapes are a little weaker than
that of an initially supercritical cloud that shows trans-sonic infall \citep{kud07}.
More quantitative analysis of the field morphological difference and the
infall speeds may distinguish the mass-to-flux ratio in a molecular cloud.

%

\subsection{Energy dissipation}
In Figure 19, we showed the time evolution of the total turbulent kinetic energy 
in isothermal subcritical clouds. It shows that the kinetic energy dissipates 
efficiently even in the flux-freezing limit ($\alpha=0$).
This is different from the result of \citet{bas10}. They found 
that the turbulent kinetic energy persists in an isothermal subcritical cloud in 
the flux-freezing limit.
They showed that the energy dissipation occurs effectively only in the
stage of the first nonlinear compression.
After that, the turbulent energy persists, and oscillates about an average value.
The average value is estimated to be about half of the initial turbulent kinetic energy 
for the same parameters as those of our fiducial model.
On the other hand, our simulation shows that energy dissipation occurs
effectively even in the flux-freezing limit; 
the energy dissipation occurs almost exponentially during several oscillations, although 
about 12 percent of the initial kinetic energy remains at the final stage.
The difference is likely due to the different vertical boundary conditions 
on the molecular cloud.
\citet{bas10} use a thin-disk approximation of the cloud, and adopt the current-free
condition of the magnetic field outside of the disk, assuming the outside density 
is negligibly small.
We study the problem by three-dimensional MHD and without a thin-disk approximation, 
but assume that the molecular cloud is surrounded by warm gas whose density 
is 10 times smaller than the molecular cloud. The computational region in
our model is $0<z<4H_0$, so does not cover a very large dynamic range of 
densities. Therefore, future three-dimensional simulations with a 
larger computational region along the $z$-direction and including low density 
gas outside of the disk may make for more revealing comparisons with 
the result of \citet{bas10}.

\subsection{Additional parameter surveys for numerical accuracy}
In addition to the parameter study in Table 1, we performed simulations of 
models with the 
same parameters as models V4 and V1, but different random realizations of the 
initial velocity fluctuations.
The core formation time for different realizations varies by
about 25\% in these samples. The overall evolution and 
the result of the accelerated core formation by nonlinear flows 
were not altered by the different initial random realizations.
We performed simulations with different spatial resolutions for the model V4. 
In the case of the highest resolution of $(N_x, N_y, N_z)=(512,512,40)$,
the core formation time is $t_{\rm core} =  14.2t_0$.
There is a slight tendency that the core formation time becomes a little shorter for
the high resolution cases \citep{kud09}.
However, we should note that a realization of the random perturbations for 
the initial velocity fluctuation is also not the same for the different resolutions, 
because the random perturbations are input on all scales down to the grid scale. 
At the least, we can say that the result of the accelerated core formation 
by nonlinear flows is not altered significantly by spatial resolution.
We also performed a simulation with a different boundary condition of magnetic fields 
for the model V4. Instead of the symmetrical boundaty at $z=z_{\rm out}$, we used
the vertical field condition for the magnetic field, 
i.e., $B_x=B_y=0$ for $z \ge z_{\rm out}$.
The overall evolution and the result of the accelerated core formation 
were not also altered by this. The core formation time for the different baounday 
varies by about 0.2\% in this sample.

\section{Summary}

We have performed fully three-dimensional magnetohydrodynamic simulations of 
collapsing core formation in molecular clouds with subcritical mass-to-flux ratio,
including ambipolar diffusion.
Some of our major findings are as follows.

\begin{itemize}
\item
Core formation in subcritical clouds is generally slow.
The core develops gradually over an ambipolar diffusion time.
When the initial mass-to-flux ratio is 0.5 times a critical value,
the formation time is about $3 \times 10^7$ years for an initial 
midplane number density $10^4$ cm$^{-3}$.

\item
The core formation time is shortened by strong velocity fluctuations.
When the average strength of the velocity fluctuation is 3 times the sound speed,
the formation time is about $5 \times 10^6$ years for the same cloud described above.

\item
The core formation time scales as $t_{core} \propto 1/\sqrt{\rho_{peak}}$, where
$\rho_{peak}$ is the value of the density peak during the first compression 
in the time evolution of the maximum density. The density dependence 
is similar to that derived by \citet{mou99}.

\item
In the case of a highly subcritical cloud, the core formation time does not
strongly depend on the initial mass-to-flux ratio even when there is strong
velocity fluctuation. This tendency is consistent with the result of 
the linear analysis of thin-disk approximation studied by \citet{cio06}.

\item
Once a core forms, the density, velocity, and magnetic field structure of the core
do not strongly depend on the initial strength of the velocity fluctuation.
The infall velocities are subsonic and the magnetic field lines show 
weak hourglass shapes.

\end{itemize}

\acknowledgments
Numerical simulations were done on the SX-9 at 
the Center for Computational Astrophysics in 
National Astronomical Observatory of Japan. 
S. B. was supported by a Discovery Grant from the Natural
Sciences and Engineering Research Council of Canada.



\clearpage

\begin{deluxetable}{cccccccll}
\tabletypesize{\scriptsize}
\tablecaption{Model and Parameters}
\tablewidth{0pt}
\tablehead{
\colhead{Model} & \colhead{$\beta_{0}$} & \colhead{$\alpha$} & \colhead{Spectrum} & \colhead{$v_a/c_s$} &
 \colhead{$t_{core}/t_0$} &  \colhead{Comments}
}
\startdata
V0 & 0.25 & 0.11 & $k^{-4}$ & 0.01 &     136   & \\
V1 & 0.25 & 0.11 & $k^{-4}$ & 0.1  &     88.9  & \\
V2 & 0.25 & 0.11 & $k^{-4}$ & 1.0  &     41.8  & \\
V3 & 0.25 & 0.11 & $k^{-4}$ & 2.0  &     24.9  & \\
V4 & 0.25 & 0.11 & $k^{-4}$ & 3.0  &     18.7  & fiducial model\\
V5 & 0.25 & 0.11 & $k^{-4}$ & 4.0  &     16.1  & \\

K0 & 0.25 & 0.11 & $k^{0}$  & 0.1  &     141   & \\
K1 & 0.25 & 0.11 & $k^{0}$  & 3.0  &     67.0  & \\
K2 & 0.25 & 0.11 & $k^{0}$  & 5.0  &     64.9  & \\
K3 & 0.25 & 0.11 & $k^{0}$  & 8.0  &     55.0  & \\
K4 & 0.25 & 0.11 & $k^{0}$  & 10.0 &     44.7  & \\

A0 & 0.25 & 0.0  & $k^{-4}$ & 3.0  &     $>268$ & no collapse\\
A1 & 0.25 & 0.05 & $k^{-4}$ & 3.0  &     63.6  & \\
A2 & 0.25 & 0.2  & $k^{-4}$ & 3.0  &     3.65  & \\
A3 & 0.25 & 0.05 & $k^{-4}$ & 0.1  &     220   & \\
A4 & 0.25 & 0.2  & $k^{-4}$ & 0.1  &     44.9  & \\

B0 & 0.04 & 0.11 & $k^{-4}$ & 3.0  &     29.9  & \\
B1 & 0.09 & 0.11 & $k^{-4}$ & 3.0  &     24.8  & \\
B2 & 0.36 & 0.11 & $k^{-4}$ & 3.0  &     6.94  & \\
B3 & 4.0 & 0.11 & $k^{-4}$ & 3.0  &      1.01  & initially supercritical \\
B4 & 9.0 & 0.11 & $k^{-4}$ & 3.0  &      0.97  & initially supercritical \\
B5 & 0.04 & 0.11 & $k^{-4}$ & 0.1  &    114   & \\
B6 & 0.09 & 0.11 & $k^{-4}$ & 0.1  &    108   & \\
B7 & 0.36 & 0.11 & $k^{-4}$ & 0.1  &    68.6  & \\
B8 & 4.0 & 0.11 & $k^{-4}$ & 0.1  &      6.93  & initially supercritical \\
B9 & 9.0 & 0.11 & $k^{-4}$ & 0.1  &      5.99  & initially supercritical \\
B10 & $\infty$ & 0.11 & $k^{-4}$ & 0.1  &      5.38  & initially supercritical \\

\enddata
\tablecomments{$\beta_{0}$ is the initial plasma $\beta$ at $z=0$. 
$\alpha$ is a dimensionless ambipolar diffusion coefficient. 
$v_a$ is the amplitude of the initial velocity fluctuation.
$t_{core}$ is the time of collapsing core formation. 
}
\end{deluxetable}

\clearpage



\begin{figure}
\epsscale{.90}
\plotone{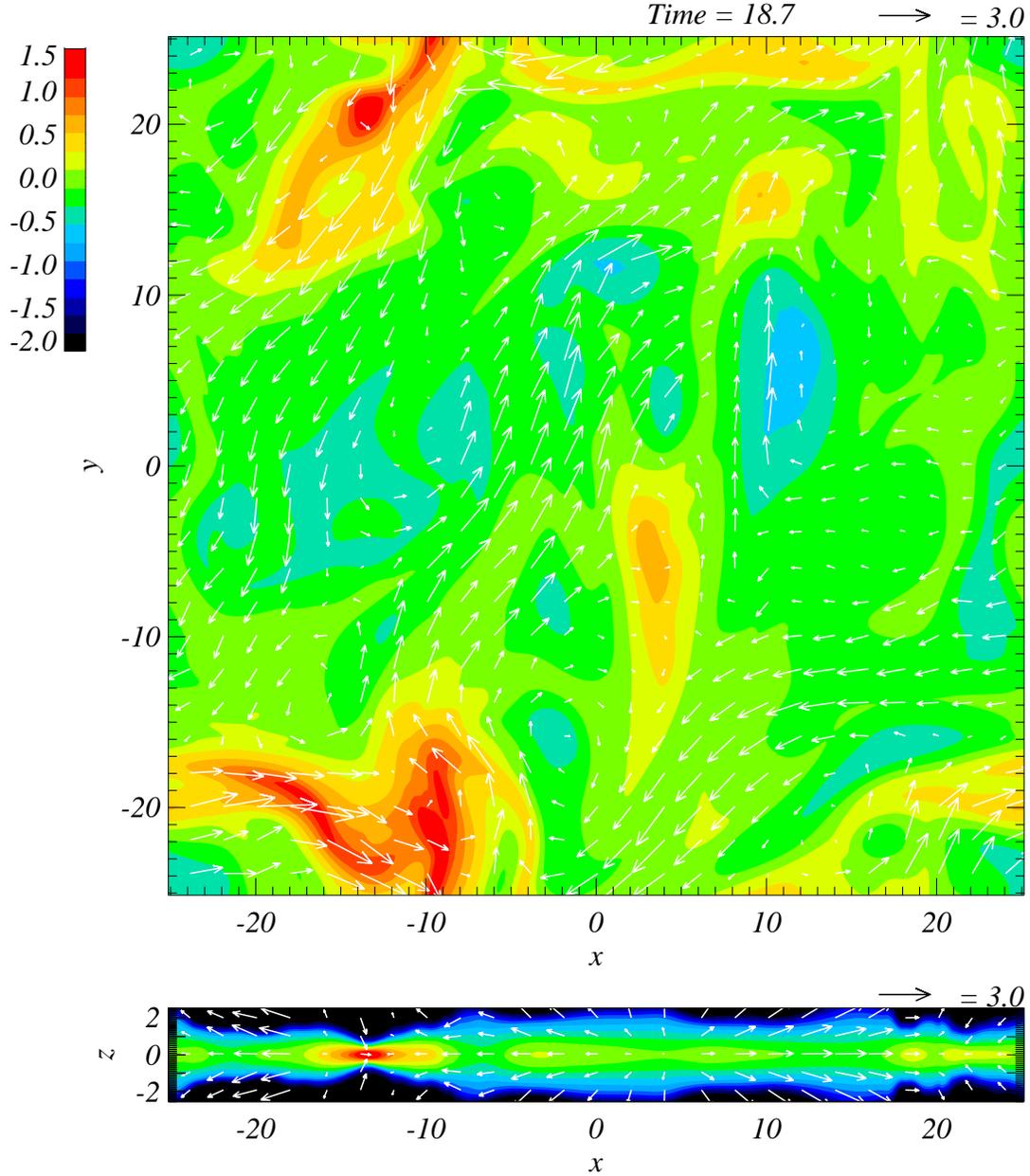}
\caption{
Logarithmic density contours at $t=18.7t_0$ for the model V4.
The model V4 is the fiducial model, in which the initial velocity amplitude $v_a
= 3\,c_{s0}$, its spectrum is $v_k^2 \propto k^{-4}$, 
the initial normalized mass-to-flux ratio is about 0.5 (i.e., $\beta_0=0.25$),
and the dimensionless ambipolar diffusion coefficient has a typical value ($\alpha=0.11$).
The unit of time $t_0$ is $\simeq  2.5 \times 10^5$ yr, and $H_0$ is
$\simeq 0.05$ pc.
The top panel shows the $x-y$ cross section at $z=0$, and the
bottom shows the $x-z$ cross section at $y=-20.7H_0$.
Arrows show velocity vectors on each cross section.
The unit of the velocity vector is three times the sound speed $c_{s0}$.
\label{fig1}}
\end{figure}

\clearpage

\begin{figure}
\epsscale{.90}
\plotone{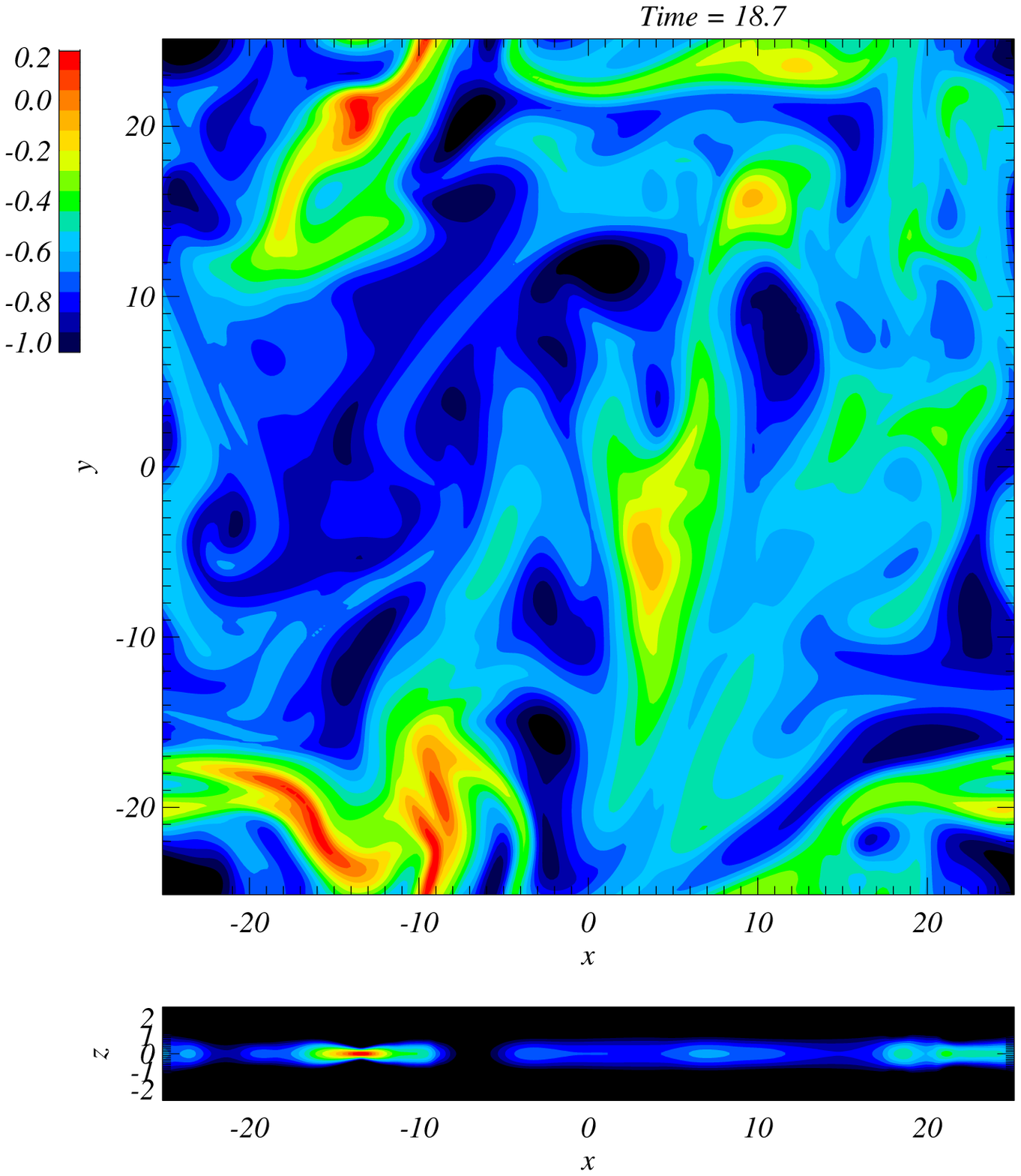}
\caption{
Logarithmic plasma $\beta$ contours at $t=18.7t_0$ for the model V4.
The top panel shows the $x-y$ cross section at $z=0$, and the
bottom panel shows the $x-z$ cross section at $y=-20.7H_0$.
\label{fig2}}
\end{figure}

\clearpage

\begin{figure}
\epsscale{1.0}
\plotone{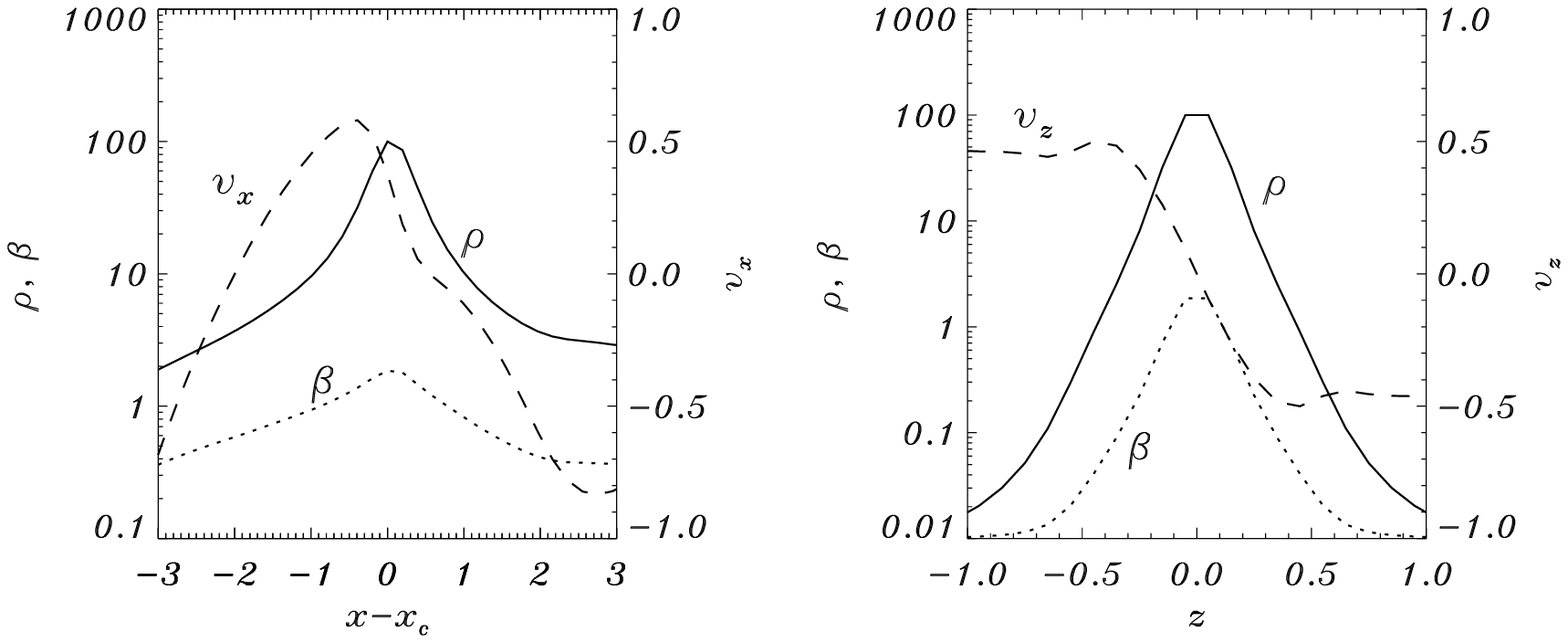}
\caption{
Left: The density (solid line), $x$-velocity (dashed line) and plasma $\beta$ (dotted line) 
along a $x$-axis cut at $y=20.7H_0$ and $z=0$ in the snapshot shown in Fig. 1 and Fig. 2.
The $x$-positions are measured by offset from $x_c=-13.4H_0$, which is the maximum density
point for the core.
Right:  The density, $z$-velocity and plasma $\beta$ along a $z$-axis 
cut at $x=-13.4H_0$ and $y=20.7H_0$ in the snapshot shown 
in Fig. 1 and Fig. 2. The line styles are the same as those in the left panel.
The unit of length $H_0$ is $\simeq 0.05$ pc.
\label{fig3}}
\end{figure}

\clearpage

\begin{figure}
\epsscale{1.0}
\plotone{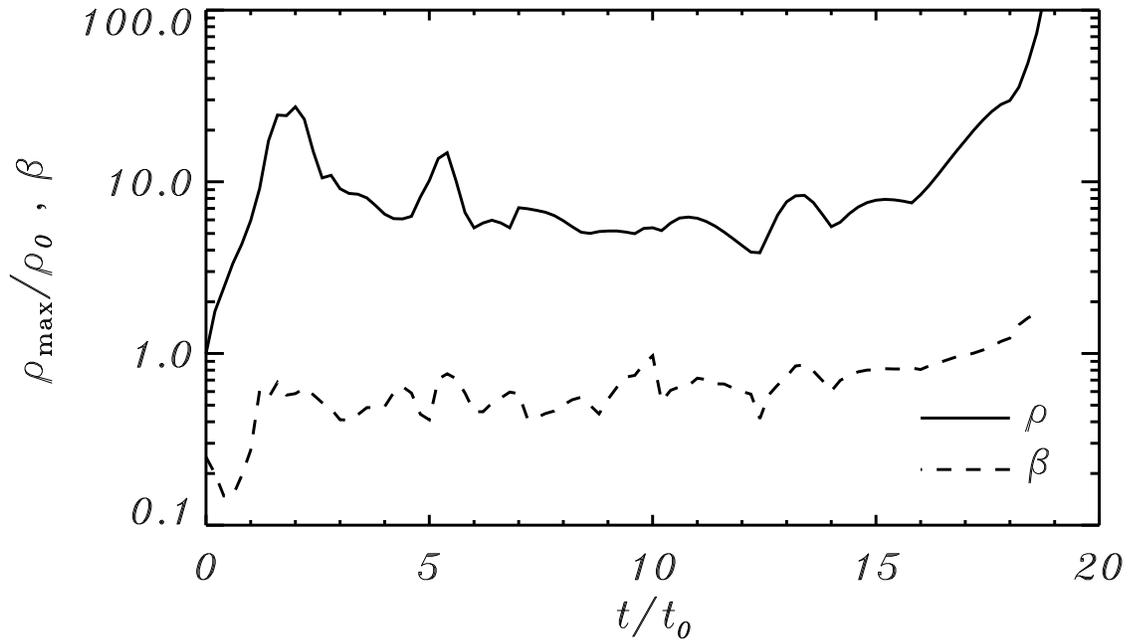}
\caption{
Time evolution of the maximum density (solid line) at $z=0$, and 
the evolution of plasma $\beta$ at the location of maximum density (dashed line)
for the model V4.
The unit of time $t_0$ is $\simeq  2.5 \times 10^5$ yr.
\label{fig4}}
\end{figure}

\clearpage

\begin{figure}
\epsscale{1.0}
\plotone{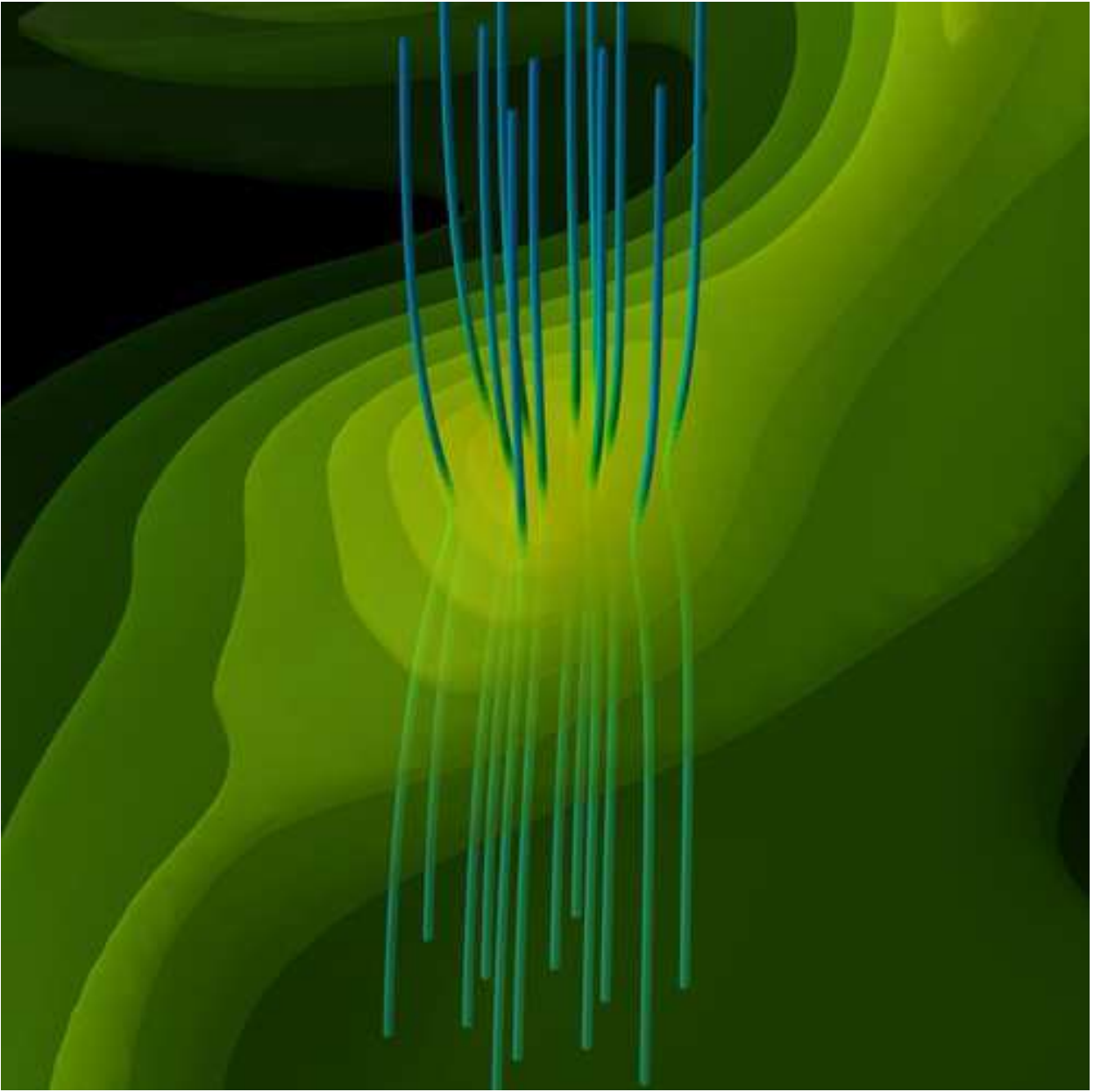}
\caption{
Isosurface of the logarithmic density, and the magnetic field lines near a core 
in the vicinity of $(x,y,z)=(-13.4H_0,20.7H_0,0.0)$ in Fig. 1. 
\label{fig5}}
\end{figure}

\clearpage

\begin{figure}
\epsscale{0.9}
\plotone{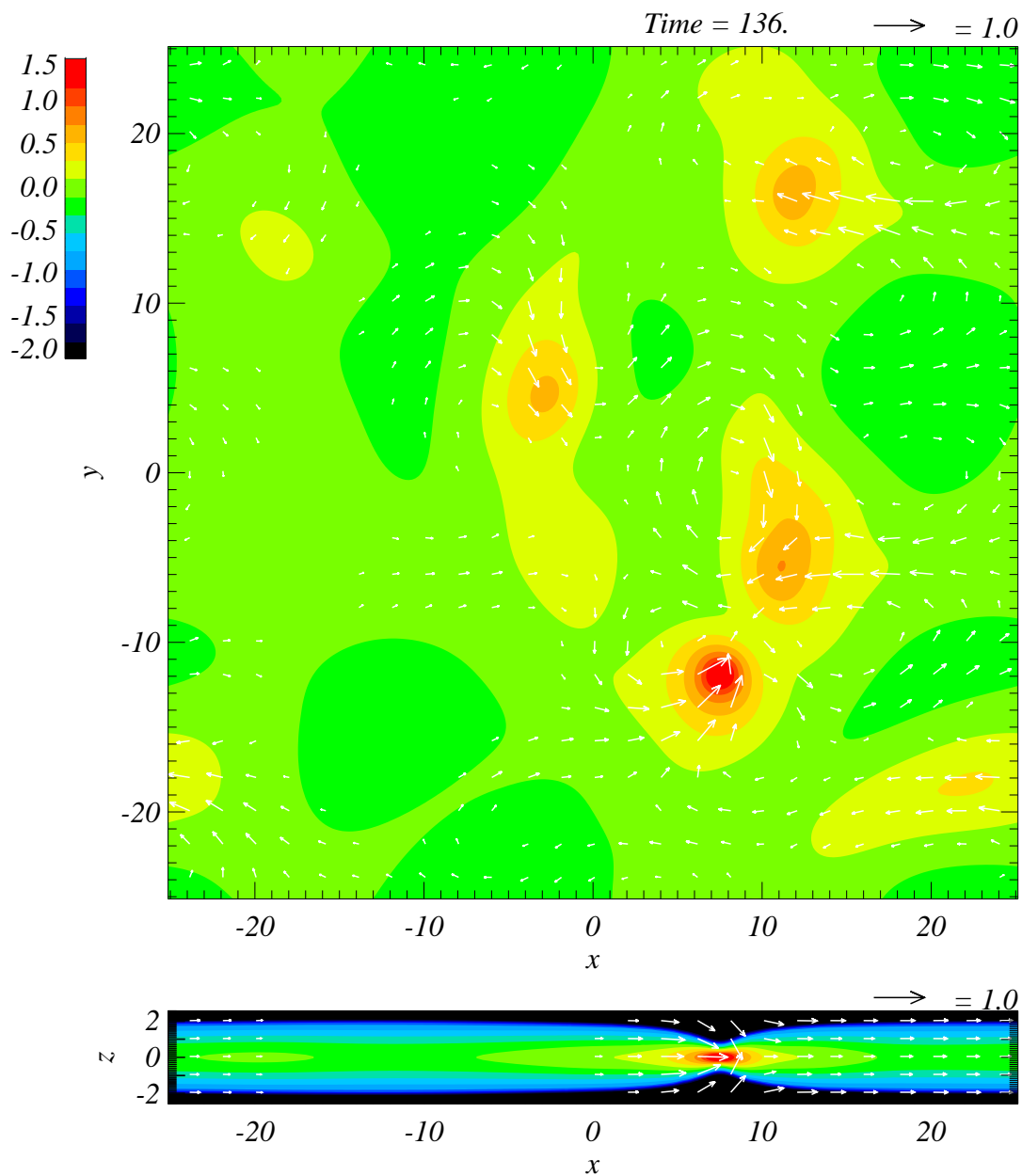}
\caption{
Logarithmic density contours at $t=136t_0$ for the model V0.
The model V0 coresponds to the model for initial velocity amplitude $v_a
= 0.01 \, c_{s0}$, but other parameters are the same as 
the fiducial model V4.
The top panel shows the $x-y$ cross section at $z=0$, and the
bottom panel shows the $x-z$ cross section at $y=-11.9H_0$.
Arrows show velocity vectors on each cross section.
The unit of the velocity vector is the sound speed $c_{s0}$.
\label{fig6}}
\end{figure}

\clearpage

\begin{figure}
\epsscale{0.9}
\plotone{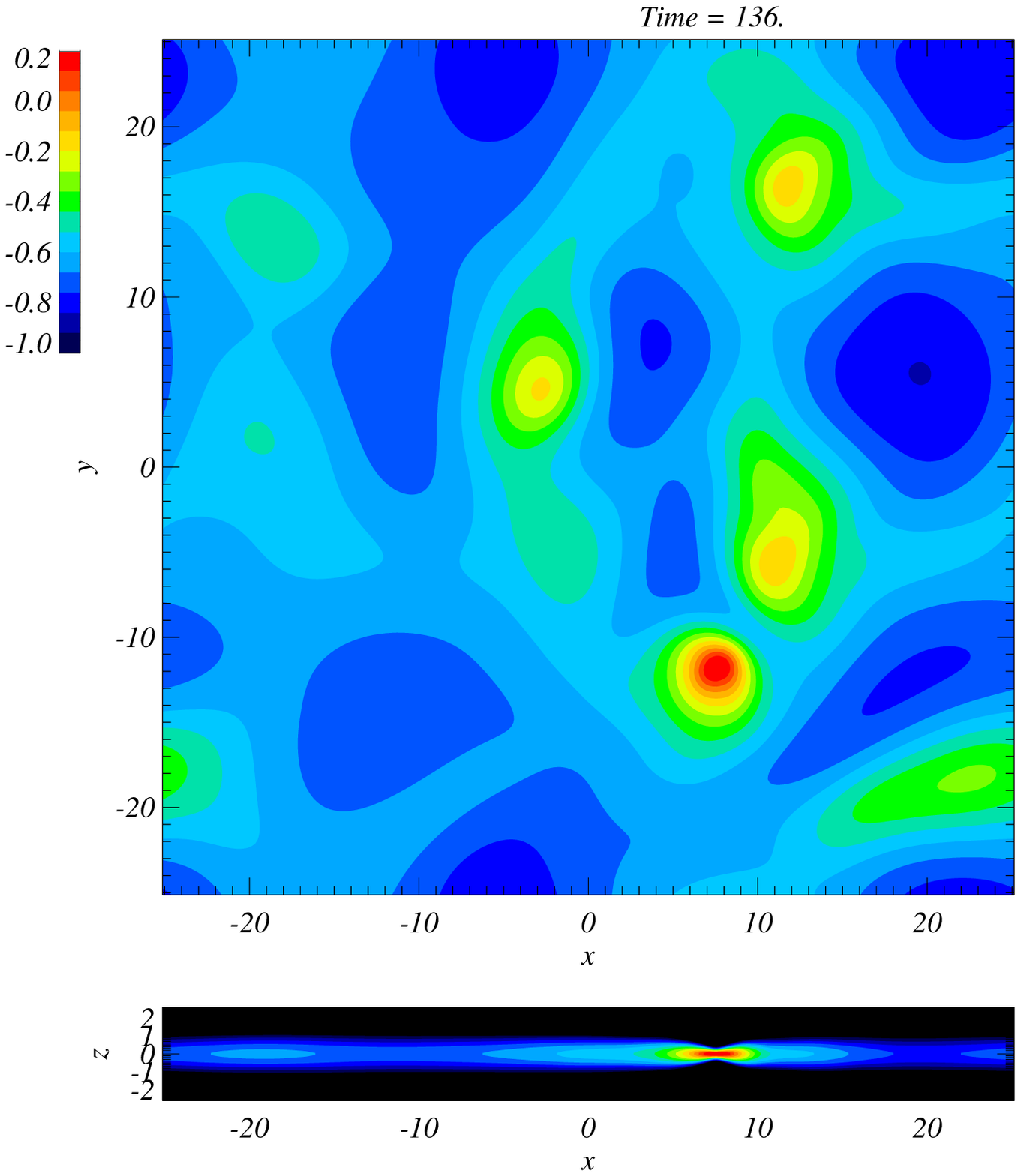}
\caption{
Logarithmic plasma $\beta$ contours at $t=136t_0$ for the model V0.
The top panel shows the $x-y$ cross section at $z=0$, and the
bottom panel shows the $x-z$ cross section at $y=-11.9H_0$.
\label{fig7}}
\end{figure}

\clearpage

\begin{figure}
\epsscale{1.0}
\plotone{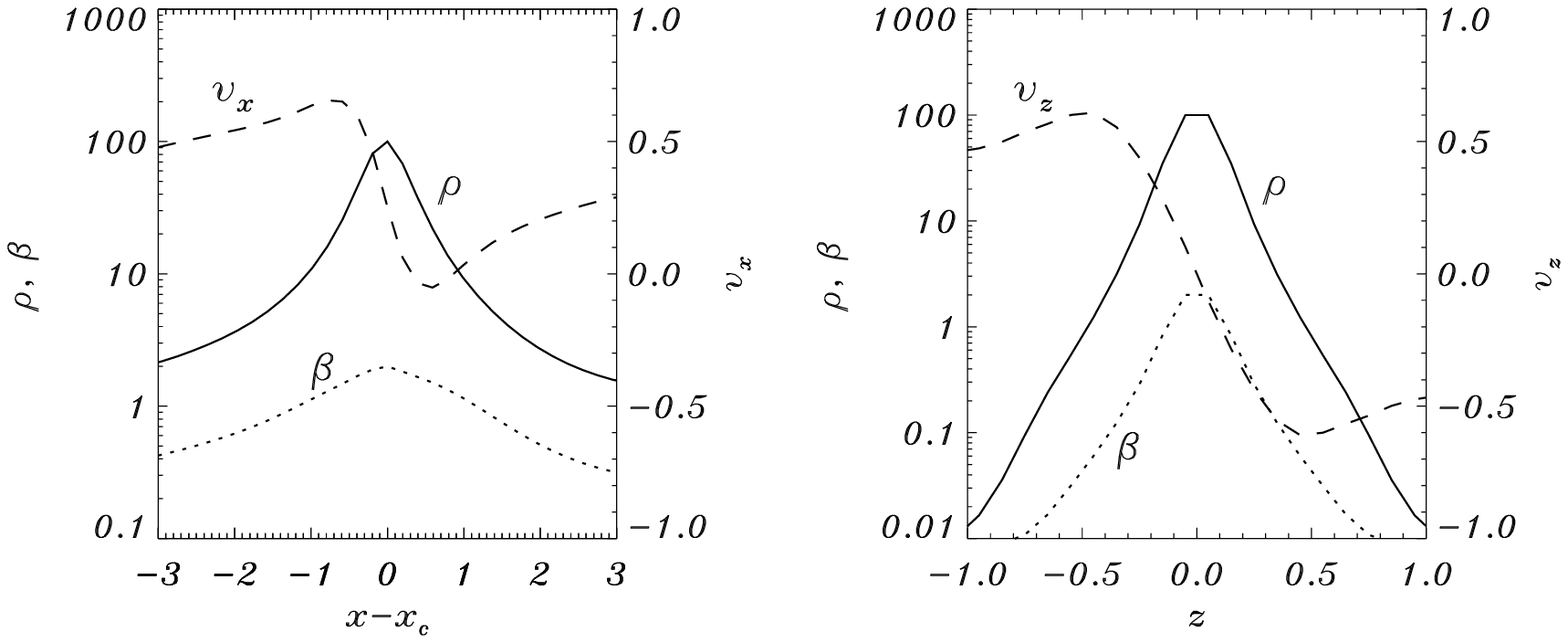}
\caption{
Left: The density (solid line), $x$-velocity (dashed line) and plasma $\beta$ (dotted line) 
along a $x$-axis cut at $y=-11.9H_0$ and $z=0$ in the snapshot shown in Fig. 6 and Fig. 7.
The $x$-positions are measured by offset from $x_c=7.6H_0$, which is the maximum density
point for the core.
Right:  The density, $z$-velocity and plasma $\beta$ along a $z$-axis 
cut at $x=7.6H_0$ and $y=-11.9H_0$ in the snapshot shown 
in Fig. 6 and Fig. 7. The line styles are the same as those in the left panel.
The unit of length $H_0$ is $\simeq 0.05$ pc.
\label{fig8}}
\end{figure}

\clearpage

\begin{figure}
\epsscale{1.0}
\plotone{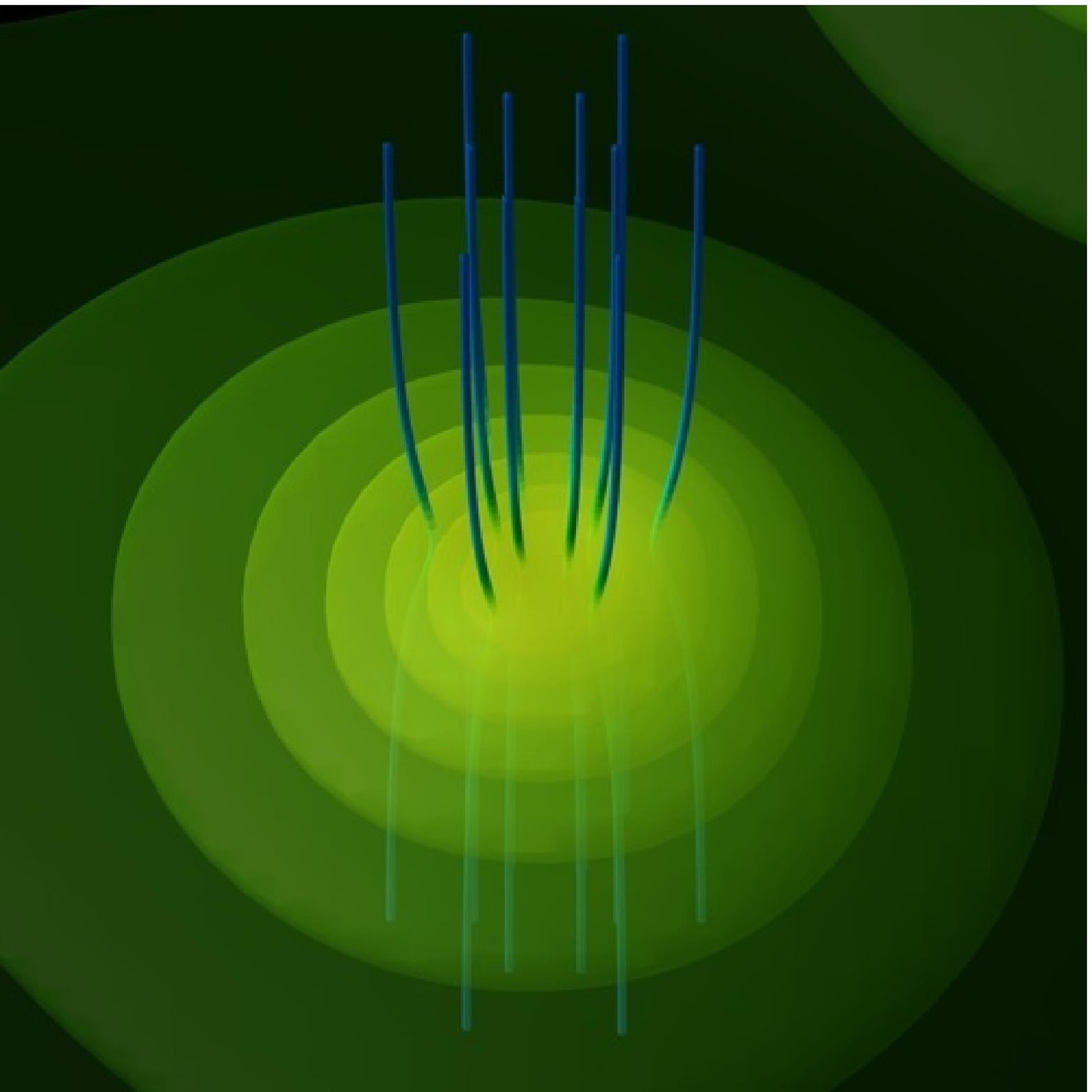}
\caption{
Isosurface of the logarithmic density and magnetic field lines near the core 
in the vicinity of $(x,y,z)=(7.6H_0,-11.9H_0,0.0)$ in Fig. 6. 
\label{fig9}}
\end{figure}

\clearpage

\begin{figure}
\epsscale{0.9}
\plotone{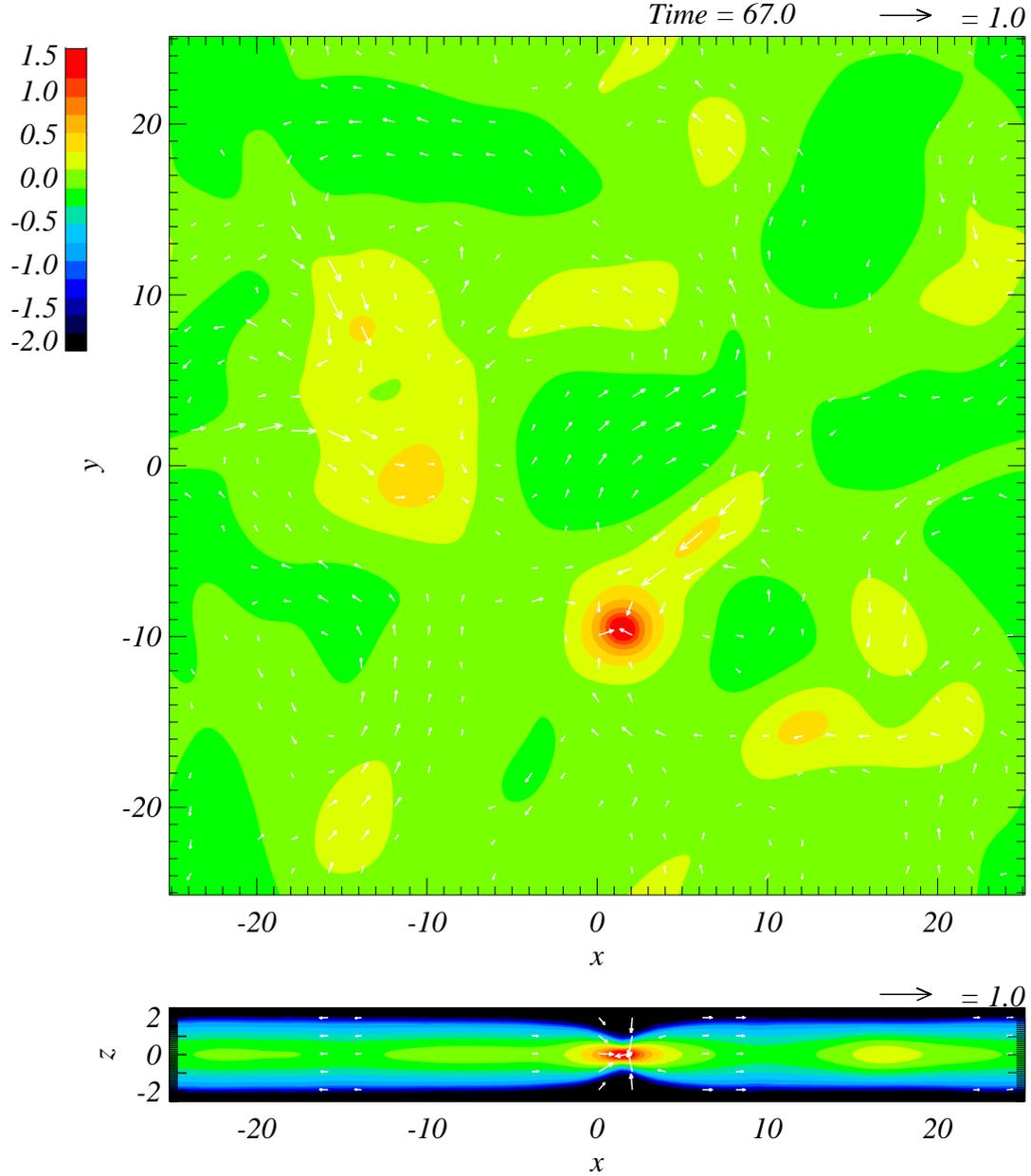}
\caption{
Logarithmic density contours at $t=67.0t_0$ for the model K1.
The model K1 has an initial velocity 
spectrum $v_k^2 \propto k^{0}$, but other parameters are the same as 
in the fiducial model V4.
The top panel shows the $x-y$ cross section at $z=0$, and the
bottom panel shows the $x-z$ cross section at $y=-9.5H_0$.
Arrows show velocity vectors on each cross section.
The unit of the velocity vector is the sound speed $c_{s0}$.
\label{fig10}}
\end{figure}

\clearpage

\begin{figure}
\epsscale{0.9}
\plotone{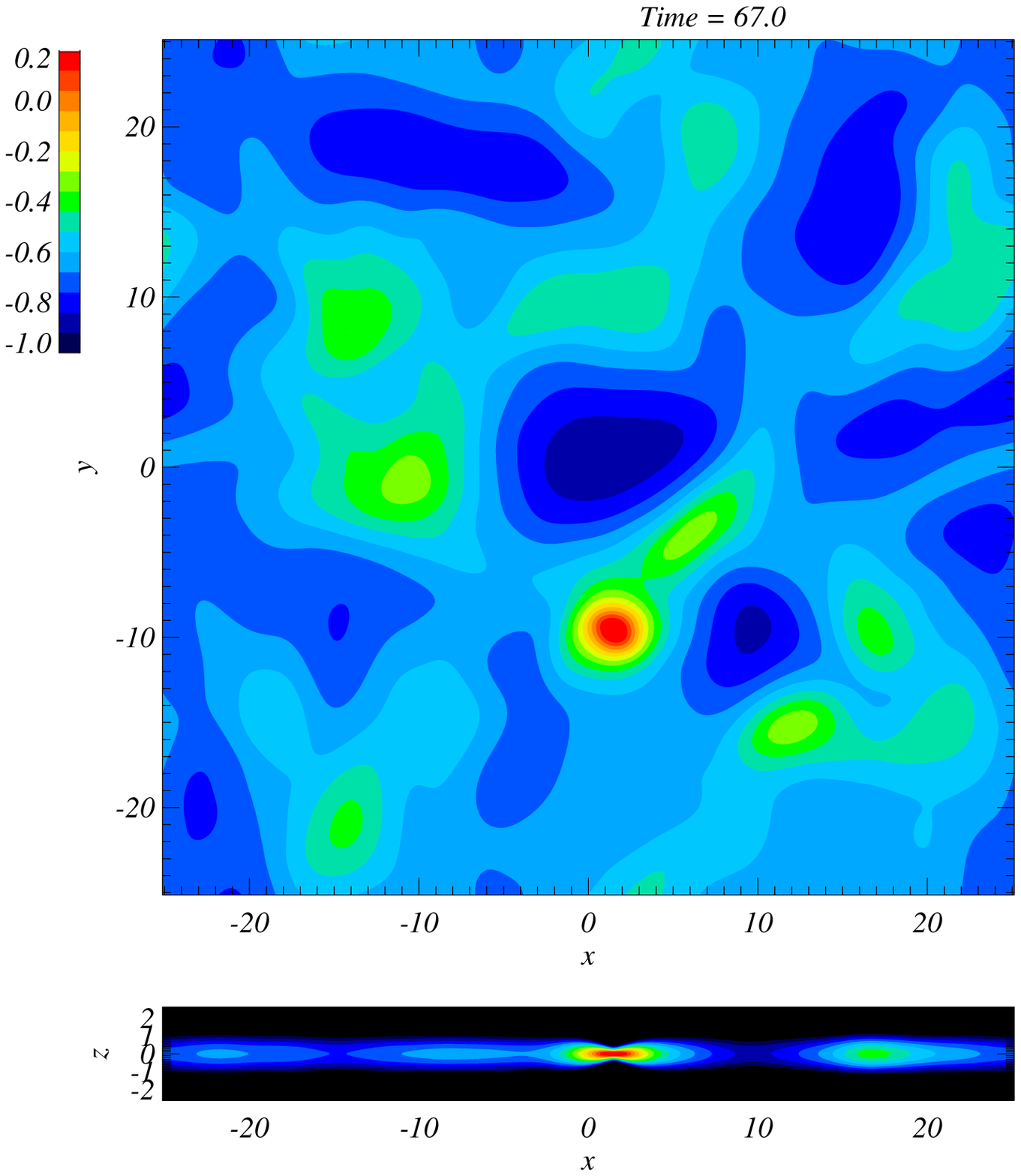}
\caption{
Logarithmic plasma $\beta$ contours at $t=67.0t_0$ for the model K1.
The top panel shows the $x-y$ cross section at $z=0$, and the
bottom panel shows the $x-z$ cross section at $y=-9.5H_0$.
\label{fig11}}
\end{figure}

\clearpage

\begin{figure}
\epsscale{1.0}
\plotone{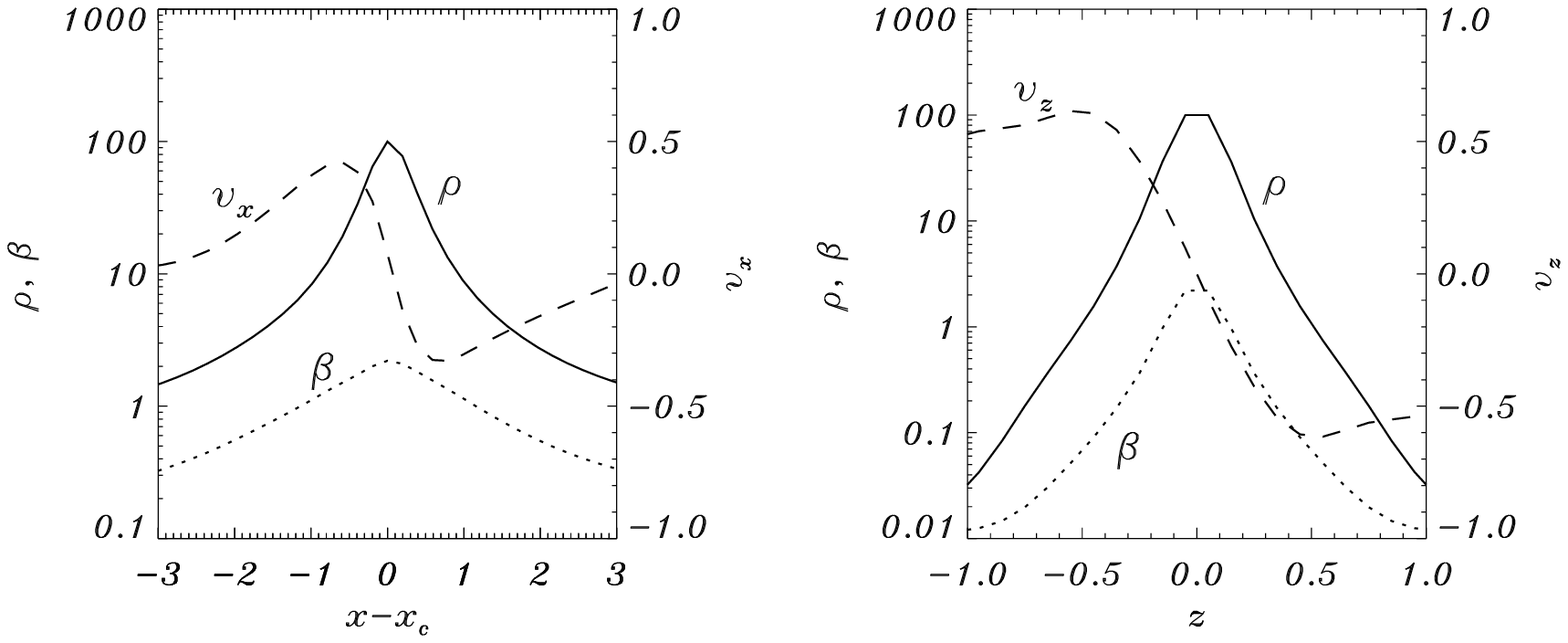}
\caption{
Left: The density (solid line), $x$-velocity (dashed line) and plasma $\beta$ (dotted line) 
along a $x$-axis cut at $y=-9.5H_0$ and $z=0$ in the snapshot shown in Fig. 10 and Fig. 11.
The $x$-positions are measured by offset from $x_c=-1.5H_0$, which is the maximum density
point for the core.
Right:  The density, $z$-velocity and plasma $\beta$ along a $z$-axis 
cut at $x=-1.5H_0$ and $y=-9.5H_0$ in the snapshot shown 
in Fig. 10 and Fig. 11. The line styles are the same as those in the left panel.
The unit of length $H_0$ is $\simeq 0.05$ pc.
\label{fig12}}
\end{figure}

\clearpage

\begin{figure}
\epsscale{1.0}
\plotone{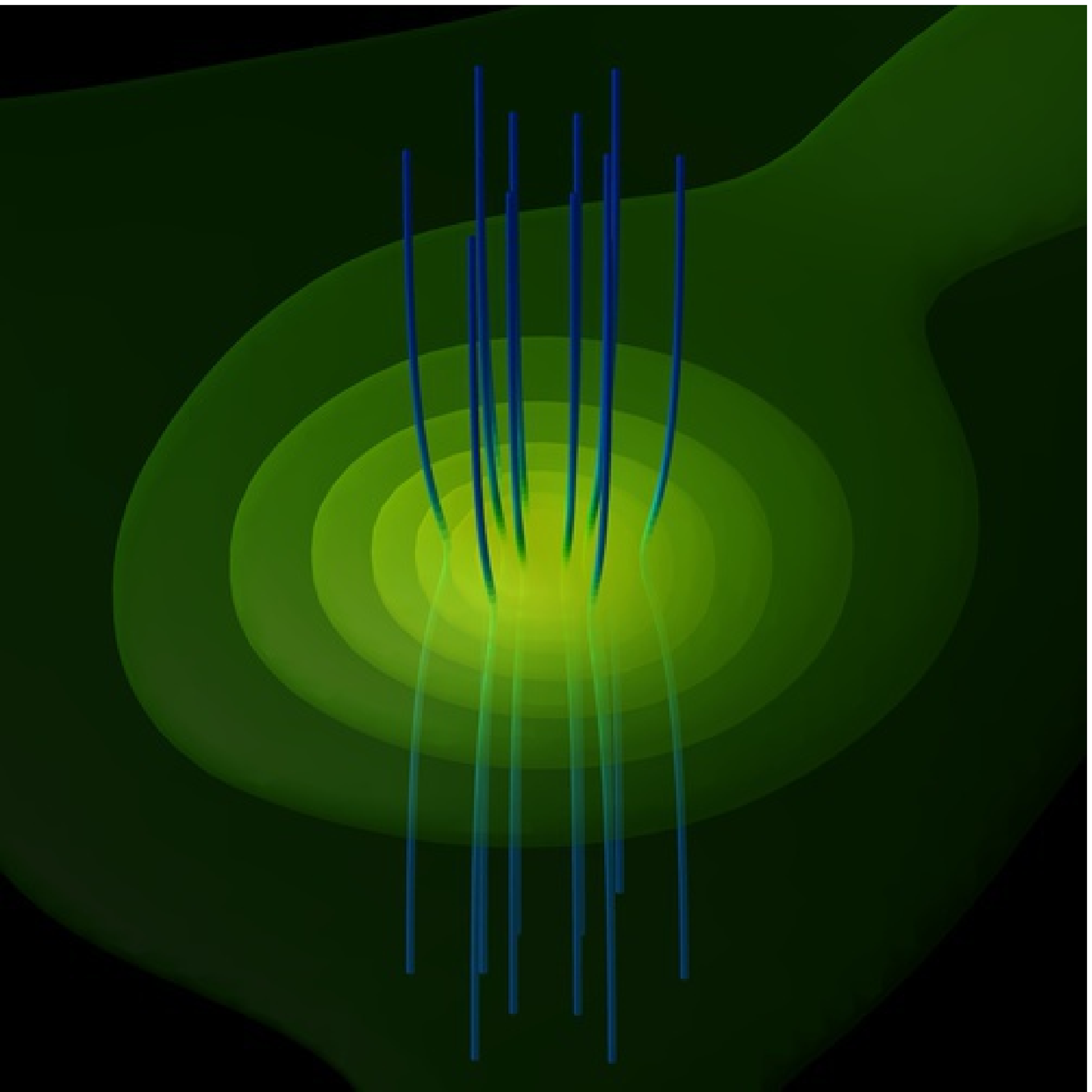}
\caption{
Isosurface of the logarithmic density and magnetic field lines near the core
in the vicinity of $(x,y,z)=(-1.5H_0,-9.5H_0,0.0)$ in Fig. 10. 
\label{fig13}}
\end{figure}

\clearpage

\begin{figure}
\epsscale{1.0}
\plotone{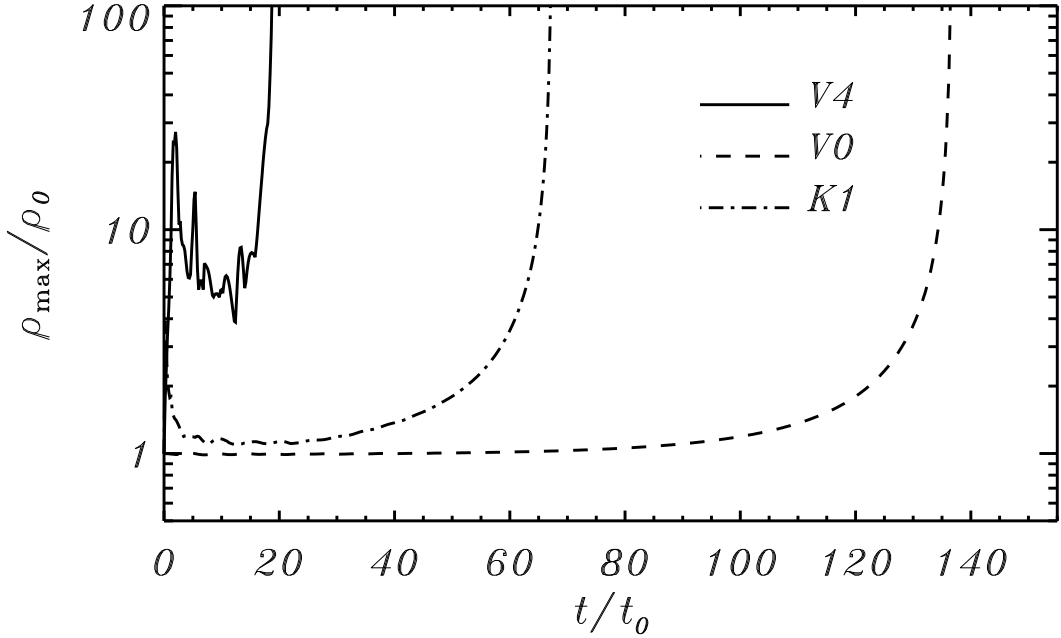}
\caption{
Time evolution of the maximum density at $z=0$
for the model V4 (solid line), V0 (dashed line), and K1 (dash-dotted line),
showing the difference between models with different velocity perturbation.
The model V4 is the fiducial model, in which $v_a
= 3\,c_{s0}$, its spectrum is $v_k^2 \propto k^{-4}$, 
the initial normalized mass-to-flux ratio is about 0.5 (i.e., $\beta_0=0.25$),
and the dimensionless ambipolar diffusion coefficient has a typical value ($\alpha=0.11$).
The model V0 has $v_a = 0.01\, c_{s0}$, but other parameters are the same as those of model V4.
The model K1 has a velocity 
spectrum $v_k^2 \propto k^{0}$, but other parameters are the same as those of model V4.
The unit of time $t_0$ is $\simeq  2.5 \times 10^5$ yr.
\label{fig14}}
\end{figure}

\clearpage

\begin{figure}
\epsscale{1.0}
\plotone{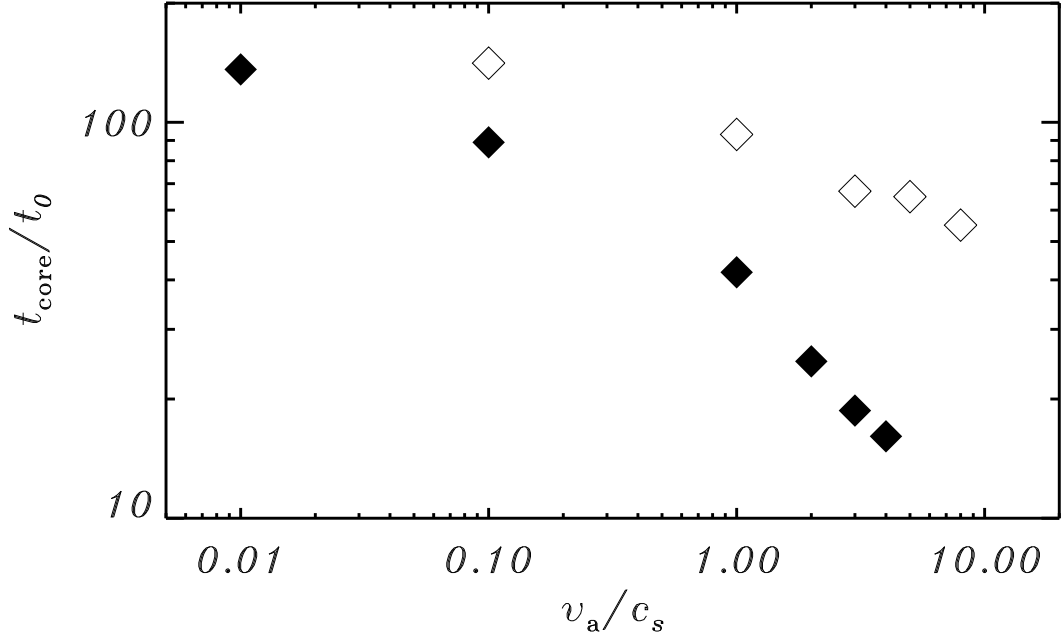}
\caption{
Core formation time as a function of the initial velocity amplitude.
The filled squares represent the results for the initial $k^{-4}$ spectrum
(model V0, V1, V2, V3, V4 and V5), and the open squares represent those
for the initial $k^{0}$ spectrum (model K0, K1, K2, K3 and K4).
The unit of time $t_0$ is $\simeq  2.5 \times 10^5$ yr.
\label{fig15}}
\end{figure}

\clearpage

\begin{figure}
\epsscale{1.0}
\plotone{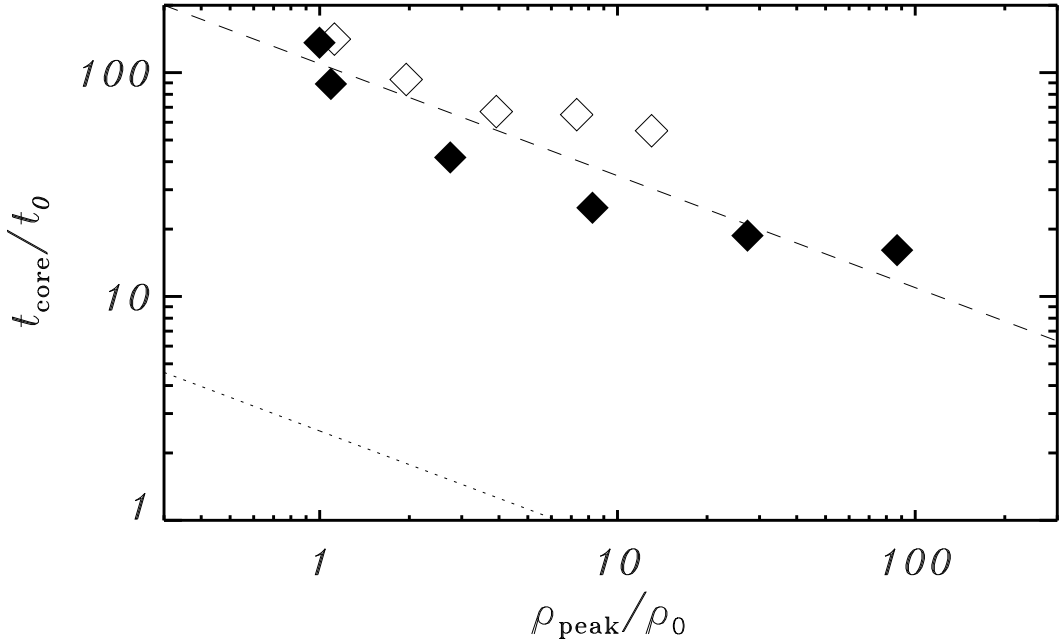}
\caption{
Core formation time as a function of the density peak during the first compression 
in its time evolution.
The filled squares represent the results for the initial $k^{-4}$ spectrum
(model V0, V1, V2, V3, V4 and V5), and the open squares represent those
for the initial $k^{0}$ spectrum (model K0, K1, K2, K3 and K4).
The dashed line shows that the core formation time is nearly proportional 
to $1/\sqrt{\rho_{peak}}$.
The dotted line represents the free fall time of gas with density $\rho_{peak}$ 
($= 1/\sqrt{G \rho_{peak}}$) for comparison.
The unit of time $t_0$ is $\simeq  2.5 \times 10^5$ yr.
\label{fig16}}
\end{figure}

\clearpage


\begin{figure}
\epsscale{1.0}
\plotone{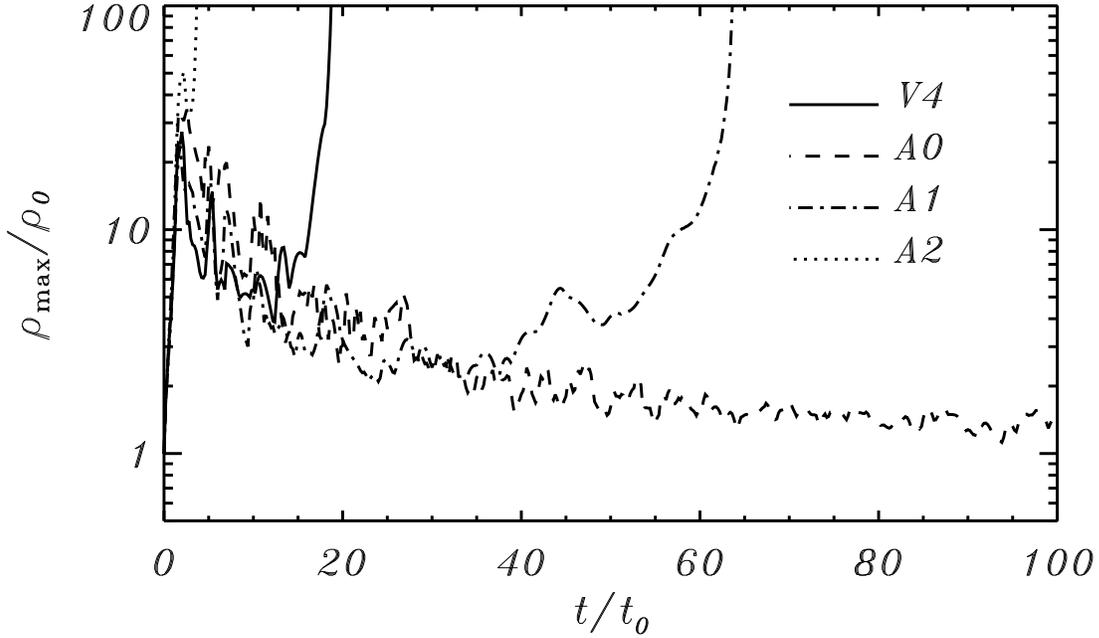}
\caption{
Time evolution of the maximum density at $z=0$
for the model V4 (solid line), A0 (dashed line), A1 (dash-dotted line)
and A2 (dotted line),
showing the difference between models with different $\alpha$.
The model V4 is the fiducial model, in which the dimensionless ambipolar 
diffusion coefficient has a typical value ($\alpha=0.11$).
The model A0, A1, and A2 correspond to the models for $\alpha=0$, $\alpha=0.05$, 
and $\alpha=0.2$, respectively, but other parameters are the same as those of model V4.
The unit of time $t_0$ is $\simeq  2.5 \times 10^5$ yr.
\label{fig17}}
\end{figure}

\clearpage

\begin{figure}
\epsscale{1.0}
\plotone{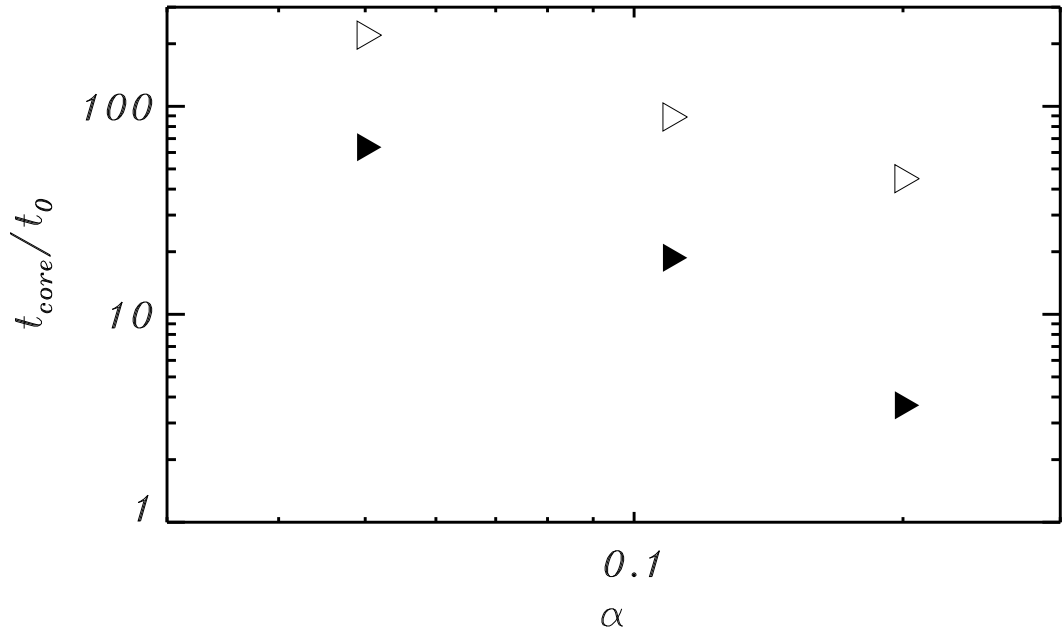}
\caption{
Core formation time as a function of $\alpha$, the dimensionless ambipolar
diffusion coefficient.
The filled triangles represent the results for an initial velocity fluctuation
$v_a=3.0c_s$ (model A1, A2 and V4), and the open triangles represent those
for an initial velocity fluctuation $v_a=0.1c_s$ (model A3, A4, and V1).
The unit of time $t_0$ is $\simeq  2.5 \times 10^5$ yr.
\label{fig18}}
\end{figure}

\clearpage

\begin{figure}
\epsscale{1.0}
\plotone{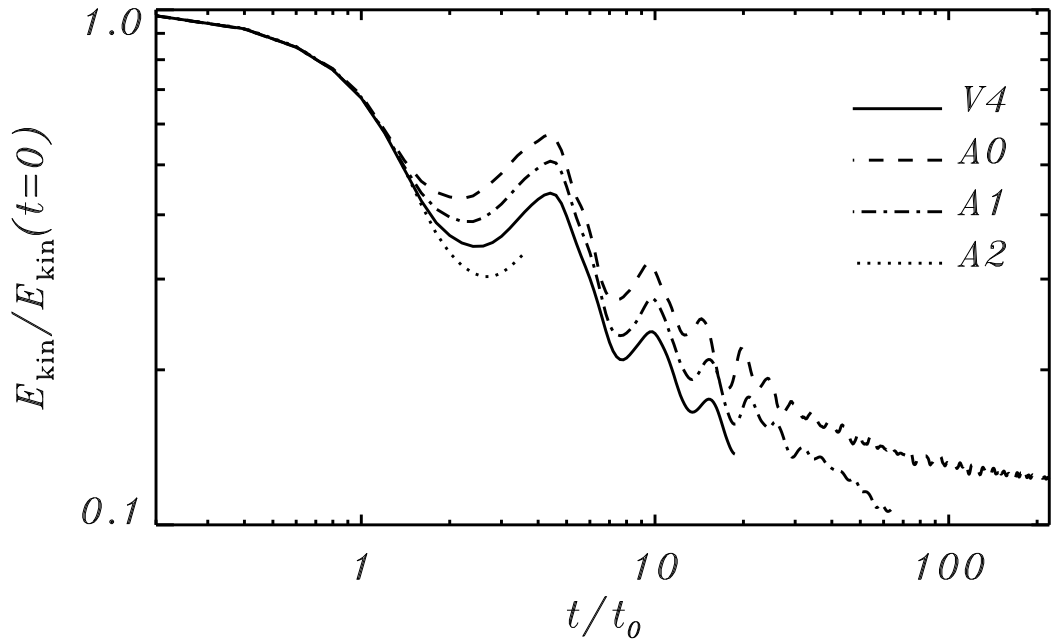}
\caption{
Time evolution of the total kinetic energies 
for the models V4 (solid line), A0 (dashed line), A1 (dash-dotted line)
and A2 (dotted line),
showing the difference between models with different $\alpha$.
The model V4 is the fiducial model, in which the dimensionless ambipolar 
diffusion coefficient has a typical value ($\alpha=0.11$).
The models A0, A1, and A2 correspond to $\alpha=0$, $\alpha=0.05$, 
and $\alpha=0.2$, respectively, but other parameters are the same as those of model V4.
The unit of time $t_0$ is $\simeq  2.5 \times 10^5$ yr.
\label{fig19}}
\end{figure}

\clearpage

\begin{figure}
\epsscale{1.0}
\plotone{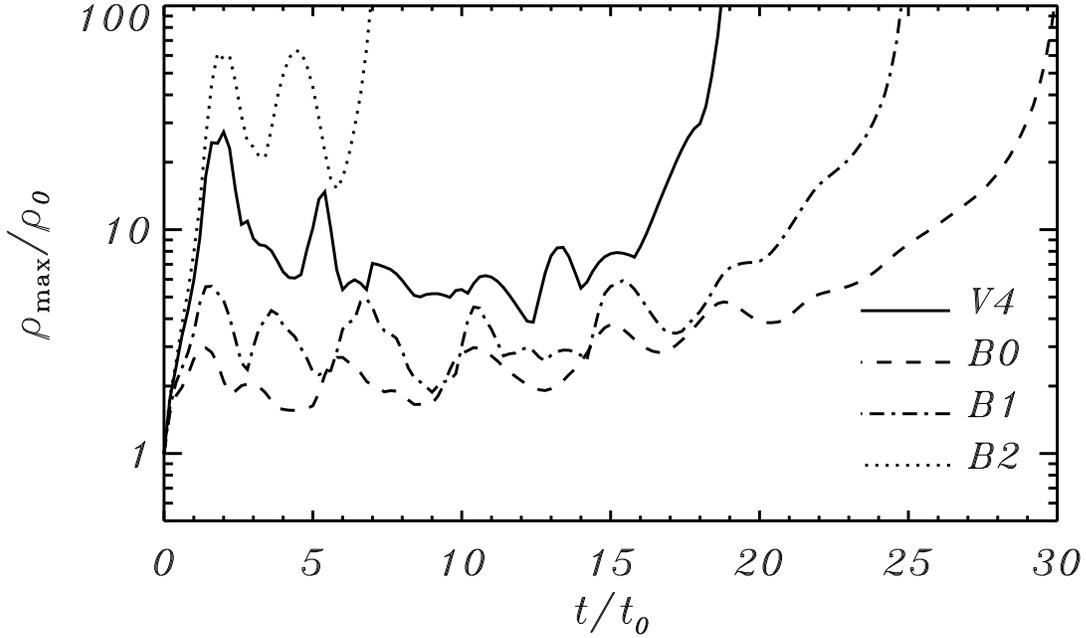}
\caption{
Time evolution of the maximum density on $z=0$
for the models V4 (solid line), B0 (dashed line), B1 (dash-dotted line)
and B2 (dotted line),
showing the difference between models with different $\beta_{0}$.
The model V4 is the fiducial model, in which 
the initial normalized mass-to-flux ratio is about 0.5 (i.e., $\beta_0=0.25$).
The models B0, B1, and B2 correspond to $\beta=0.04$, $\beta=0.09$, 
and $\beta=0.36$, respectively, but other parameters are the same as those of model V4.
The unit of time $t_0$ is $\simeq  2.5 \times 10^5$ yr.
\label{fig20}}
\end{figure}

\clearpage

\begin{figure}
\epsscale{1.0}
\plotone{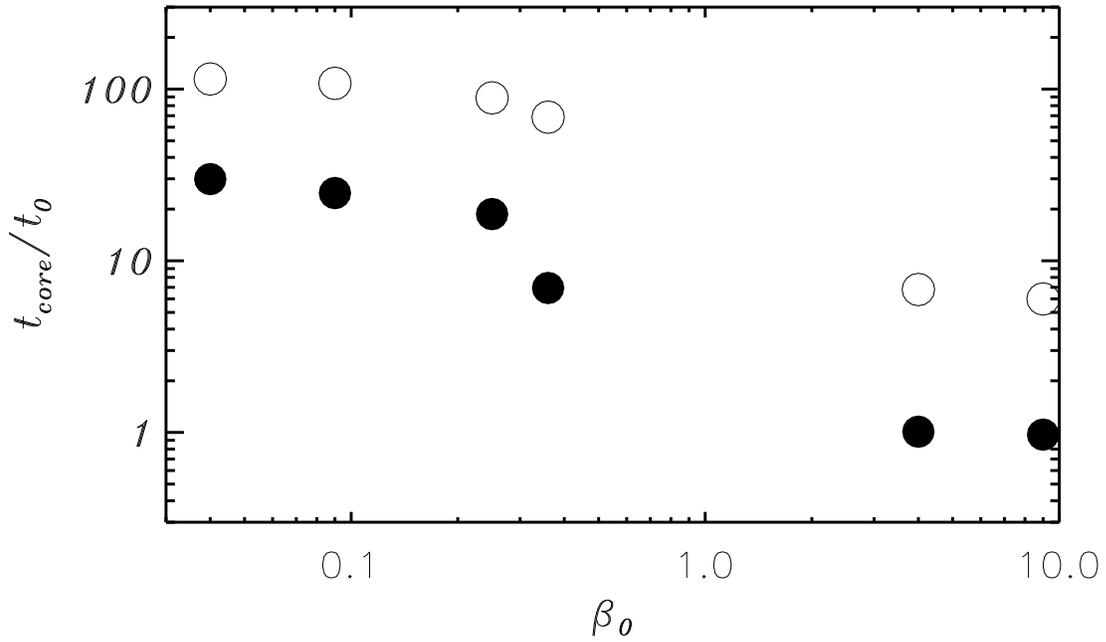}
\caption{
Core formation time as a function of $\beta_{0}$, the initial plasma $\beta$,
at $z=0$.
The filled circles represent the results for an initial velocity fluctuation
$v_a=3.0\,c_s$ (models B0, B1, B2, B3, B4 and V4), and the open circles 
represent those for an initial velocity fluctuation 
$v_a=0.1c_s$ (models B5, B6, B7, B8, B9 and V1).
The unit of time $t_0$ is $\simeq  2.5 \times 10^5$ yr.
\label{fig21}}
\end{figure}

\end{document}